%% file: main.tex
\documentclass[lettersize,journal]{IEEEtran}
\let\labelindent\relax
\usepackage{enumitem}
\usepackage{listings}
\usepackage{xcolor}

\usepackage{amsmath}
\usepackage{multirow}
\usepackage{xspace}
\usepackage{balance}
\usepackage[many]{tcolorbox}
\usepackage{booktabs}
\usepackage{threeparttable}
\usepackage[ruled,lined,linesnumbered,vlined,algo2e]{algorithm2e}
\usepackage{url}
\usepackage[noend]{algpseudocode} 
\usepackage{subfig}
\usepackage{pifont} 
\usepackage{colortbl}
\usepackage{hyperref} 
\usepackage{cleveref} 

\definecolor{OliveGreen}{rgb}{0,0.6,0} 
\newcommand{\tool}{\textsc{ACFix}\xspace} 
\newcommand{\name}{\tool}

\newcommand{\myfig}{Fig.\xspace}
\newcommand{\mysec}{\S}

\newcommand{\rev}[1]{\textcolor{black}{#1}}
\newcommand{\minor}[1]{\textcolor{black}{#1}}

\newcommand{\rcg}{\text{ACG}\xspace}
\newcommand{\gen}{\text{\textit{generator}}\xspace}
\newcommand{\val}{\text{\textit{validator}}\xspace}
\newcommand{\bla}{\text{\textit{Vanilla GPT-4}}\xspace}
\newcommand{\blb}{\text{\textit{W/o~Validator}}\xspace}
\newcommand{\blc}{\text{\textit{W/o~\rcg}}\xspace}
\newcommand{\bld}{\text{\textit{W/o~RBAC}}\xspace}

\begin{document}

\title{\name: Guiding LLMs with Mined Common RBAC Practices for Context-Aware Repair of Access Control Vulnerabilities in Smart Contracts}

\author{Lyuye Zhang, Kaixuan Li, Kairan Sun, Daoyuan Wu, Ye Liu, Haoye Tian, Yang Liu
\thanks{Lyuye Zhang, Kaixuan Li (Equal Contribution to the first author), Kairan Sun, and Yang Liu are with the College of Computing and Data Science, Nanyang Technological University, Singapore, Singapore.}
\thanks{Daoyuan Wu (Corresponding Author; \texttt{daoyuan@cse.ust.hk}) is with Department of Computer Science and Engineering, The Hong Kong University of Science and Technology, Hong Kong SAR, China. Work done while at NTU.}
\thanks{Ye Liu is with Singapore Management University. Work done while at NTU.}
\thanks{Haoye Tian is with University of Luxembourg.}
}

\markboth{Journal of \LaTeX\ Class Files,~Vol.~14, No.~8, August~2021}%
{Shell \MakeLowercase{\textit{et al.}}: A Sample Article Using IEEEtran.cls for IEEE Journals}

\IEEEpubid{0000--0000/00\$00.00~\copyright~2021 IEEE}

\maketitle

\input{rev_tex_2/0-abstract}

 \begin{IEEEkeywords}
Smart Contract, Software Security, Program Repair.
\end{IEEEkeywords}

\input{rev_tex_2/solidity-highlighting.tex}

\input{rev_tex_2/1-introduction}

\input{rev_tex_2/2-motivation}

\input{rev_tex_2/a-overview}
\input{rev_tex_2/3-prestudy}

\input{rev_tex_2/4-methodology}

\input{rev_tex_2/b-patch}
\input{rev_tex_2/5-evaluation}

\input{rev_tex_2/6-humanstudy}

\input{rev_tex_2/7-discussion}

\input{rev_tex_2/8-relatedwork}
\input{rev_tex_2/9-future_works}
\input{rev_tex_2/10-conclusion}

\bibliographystyle{IEEEtran}
\bibliography{sc_cites}
\vfill

\end{document}

%% file: rev_tex_2/0-abstract.tex
\begin{abstract}
Smart contracts are susceptible to various security issues, among which access control (AC) vulnerabilities are particularly critical.
While existing research has proposed multiple detection tools, automatic and appropriate repair of AC vulnerabilities in smart contracts remains a challenge.
Unlike commonly supported vulnerability types by existing repair tools, such as reentrancy, which are usually fixed by template-based approaches, the main obstacle of repairing AC vulnerabilities lies in identifying the appropriate roles or permissions amid a long list of non-AC-related source code to generate proper patch code, a task that demands human-level intelligence.

In this paper,
we employ the state-of-the-art GPT-4 model and enhance it with a novel approach called \tool.
The key insight is that we can mine common AC practices for major categories of code functionality and use them to guide LLMs in fixing code with similar functionality.
To this end, \tool involves offline and online phases.
In the offline phase, \tool mines a taxonomy of common Role-based Access Control practices from 344,251 on-chain contracts, categorizing 49 role-permission pairs from the top 1,000 unique samples. 
In the online phase, \tool tracks AC-related elements across the contract and uses this context information along with a Chain-of-Thought pipeline to guide LLMs in identifying the most appropriate role-permission pair for the subject contract and subsequently generating a suitable patch. 
To evaluate \tool, we built the first benchmark dataset of 118 real-world AC vulnerabilities, and our evaluation revealed that \tool successfully repaired $94.92\%$ of them, a major improvement compared to the baseline GPT-4 at only $52.54\%$.
We also conducted a human study to understand the value of \tool's repairs and their differences from human repairs.
\end{abstract}

%% file: rev_tex_2/solidity-highlighting.tex

\definecolor{verylightgray}{rgb}{.97,.97,.97}
\definecolor{codepurple}{rgb}{0.58,0,0.82}

\lstdefinelanguage{Solidity}{
	keywords=[1]{anonymous, assembly, assert, balance, break, call, callcode, case, catch, class, constant, continue, constructor, contract, debugger, default, delegatecall, delete, do, else, emit, event, experimental, export, external, false, finally, for, function, gas, if, implements, import, in, indexed, instanceof, interface, internal, is, length, library, log0, log1, log2, log3, log4, memory, modifier, new, payable, pragma, private, protected, public, pure, push, require, return, returns, revert, selfdestruct, send, solidity, storage, struct, suicide, super, switch, then, this, throw, transfer, true, try, typeof, using, value, view, while, with, addmod, ecrecover, keccak256, mulmod, ripemd160, sha256, sha3}, 
	keywordstyle=[1]\color{blue}\bfseries,
	keywords=[2]{address, bool, byte, bytes, bytes1, bytes2, bytes3, bytes4, bytes5, bytes6, bytes7, bytes8, bytes9, bytes10, bytes11, bytes12, bytes13, bytes14, bytes15, bytes16, bytes17, bytes18, bytes19, bytes20, bytes21, bytes22, bytes23, bytes24, bytes25, bytes26, bytes27, bytes28, bytes29, bytes30, bytes31, bytes32, enum, int, int8, int16, int24, int32, int40, int48, int56, int64, int72, int80, int88, int96, int104, int112, int120, int128, int136, int144, int152, int160, int168, int176, int184, int192, int200, int208, int216, int224, int232, int240, int248, int256, mapping, string, uint, uint8, uint16, uint24, uint32, uint40, uint48, uint56, uint64, uint72, uint80, uint88, uint96, uint104, uint112, uint120, uint128, uint136, uint144, uint152, uint160, uint168, uint176, uint184, uint192, uint200, uint208, uint216, uint224, uint232, uint240, uint248, uint256, var, void, ether, finney, szabo, wei, days, hours, minutes, seconds, weeks, years},	
	keywordstyle=[2]\color{codepurple}\bfseries,
	keywords=[3]{block, blockhash, coinbase, difficulty, gaslimit, number, timestamp, msg, data, gas, sender, sig, value, now, tx, gasprice, origin},	
	keywordstyle=[3]\color{violet}\bfseries,
	identifierstyle=\color{black},
	sensitive=true,
	comment=[l]{//},
	morecomment=[s]{/*}{*/},
	commentstyle=\color{gray}\ttfamily,
	stringstyle=\color{red}\ttfamily,
	morestring=[b]',
	morestring=[b]",
    xleftmargin=0.05\linewidth, 
    xrightmargin=0.0\linewidth,
}

\lstset{
	language=Solidity,
	backgroundcolor=\color{verylightgray},
	extendedchars=true,
	basicstyle=\footnotesize\ttfamily,
	showstringspaces=false,
	showspaces=false,
	numbers=left,
	numberstyle=\footnotesize,
	numbersep=9pt,
	tabsize=2,
	breaklines=true,
	showtabs=false,
	captionpos=b
}

%% file: rev_tex_2/1-introduction.tex
\section{Introduction}

Smart contracts, Turing-complete programs executed on blockchain ledgers, implement predefined programmatic logic through transaction-based invocation~\cite{wood2014ethereum}.
With the emergence of decentralized applications such as DeFi~\cite{defi} and NFTs~\cite{nft}, the use of smart contracts, especially those written in Solidity~\cite{solidity} on the Ethereum blockchain~\cite{wood2014ethereum}, has significantly expanded within the blockchain ecosystem.
Nevertheless, these contracts can be susceptible to various security vulnerabilities, including reentrancy~\cite{zheng2023turn}, integer overflow~\cite{kalra2018zeus}, front-running~\cite{daian2020flash}, price manipulation~\cite{wu2021defiranger},  etc.
Among these, Access Control (AC) vulnerabilities~\cite{ghaleb2023achecker} are particularly critical because they directly expose privileged operations to attackers, such as taking over the ownership of the contract or minting more tokens, which often lead to tremendous financial loss, e.g. an infamous attack, Parity~\cite{parity}.
\IEEEpubidadjcol

\rev{Considering the severe implications associated with access control (AC) vulnerabilities, several automated detection tools have been recently introduced to mitigate these risks, such as Ethainter~\cite{brent2020ethainter}, SPCon~\cite{liu2022finding}, AChecker~\cite{ghaleb2023achecker}, and SoMo~\cite{fang2023beyond}. Among these tools, SPCon distinguishes itself by analyzing historical transactions to infer AC policies. In contrast, the other approaches primarily employ taint analysis techniques to trace critical instructions (e.g., \texttt{selfdestruct}) or state variables (e.g., \texttt{owner}), thereby identifying potential scenarios where unauthorized parties might gain access. While these works have thoroughly addressed the detection of AC vulnerabilities, they have not provided concrete guidance or recommendations on how to remediate these issues. Furthermore, although the tools can identify potential vulnerabilities effectively, they lack an explainable reasoning process to justify or clarify the rationale behind their detections.}

While detecting AC vulnerabilities has certain information flow patterns, repairing them needs a step further to identify appropriate roles or permissions.
As a result, although numerous repair tools for smart contracts have been proposed~\cite{so2023smartfix,nguyen2021sguard,zhang2020smartshield,yu2020smart,Jin2022aroc,ferreira2022elysium,rodler2021evmpatch}, only a few of them support AC vulnerability repairs.
\rev{Unfortunately, although certain repair systems—such as Elysium~\cite{ferreira2022elysium} and SmartFix~\cite{so2023smartfix}—explicitly claim support for addressing access control (AC) vulnerabilities, their scope is restricted to fixing only a limited set of common unauthorized operations, such as \textit{Re-initialization}~\cite{init}, \textit{Suicidal}~\cite{nikolic2018finding}, and \textit{Low-level Call}~\cite{lowlevel}. However, these predefined AC misuse patterns are insufficient to comprehensively cover the complexity and diversity encountered in real-world smart contract implementations, leading to the inability of successful patch generation.}

\rev{Existing vulnerability repair tools are often constrained by predefined access control restrictions specific to the contract's owner, rendering template-based repair approaches potentially adequate for standard operational scenarios. However, the current methodological landscape presents a significant limitation in addressing unauthorized privilege escalation across broader contexts, particularly in more complex AC vulnerabilities that require nuanced automated repair mechanisms.}
For instance, the motivating example presented in \mysec\ref{sec:motivate} illustrates that an unprotected \texttt{deposit} function can also lead to unforeseen financial losses for smart contracts.
This privilege should be granted to the role \texttt{Bank} rather than the contract's owner for more flexibility.

In general, automatically and appropriately repairing AC vulnerabilities in smart contracts requires human-level intelligence.
This is because AC policies in smart contracts are commonly enforced through the Role-Based Access Control (RBAC)~\cite{sandhu1998role} mechanism, which requires setting appropriate RBAC \textit{roles} that align with corresponding privileged operations (referred to as \textit{permissions} in RBAC terminology).
Intuitively, for a repair system to function effectively, it must \textit{(i)} first achieve a human-level understanding of the functionality embedded within the vulnerable code, \textit{(ii)} then recognize appropriate RBAC roles based on this understanding, and \textit{(iii)} finally generate correct patches.
Although recent advancements in large language models (LLMs)~\cite{touvron2023llama, openai2023gpt4} allow us to utilize state-of-the-art (SOTA) models like GPT-4~\cite{openai2023gpt4}, accomplishing these three tasks still presents challenges.

Specifically, 
For\textbf{ task \textit{(i)}}, determining AC-related operations from the raw code corpus is even hard for GPT-4, given the substantial noise present within the source code. Compounding this challenge, LLMs are known to have limited attention spans, leading to a loss of focus~\cite{tian2023chatgpt}. To address this issue, we have developed a static slicing algorithm to extract the relevant code context, allowing GPT-4 to focus on it. 
    For \textbf{task \textit{(ii)}}, off-the-shelf LLMs were not inherently trained to recognize RBAC roles and their typical privileged operations, i.e., the mapping of role-permission pairs. Moreover, LLM hallucination~\cite{gpthall} could lead to unreliable output. Hence, it becomes essential to build an RBAC taxonomy, derived from common RBAC practices in smart contracts, for the LLM to select from.
    For \textbf{task \textit{(iii)}}, the patches generated might conflict with pre-existing, inaccurately implemented RBAC mechanisms. Therefore, besides building new RBAC from scratch, we also mine existing RBAC mechanisms from the source code and reuse them in the generated patches. Our evaluation suggests that this strategy is effective for addressing inadequately implemented RBAC.
   Another issue for task \textit{(iii)} is that LLMs' randomness could still occasionally divert the LLM from generating correct patches. To address this, we implemented a Multi-Agent Debate (MAD) mechanism~\cite{liang2023encouraging} to establish a loop between \textit{generator} and \textit{validator}. With such validation, \textit{validator} can effectively suppress \textit{generator}'s hallucination and ensure the generation of proper patches.

Based on the observations above, we propose a novel approach named \tool to enhance the capabilities of the state-of-the-art GPT-4 model in repairing AC vulnerabilities in smart contracts.
The key insight is that we can mine common AC practices from major categories of code functionality and use these practices to guide LLMs in fixing code with similar functionality.
Specifically, \name first conducts offline mining of common RBAC practices from 344,251 on-chain contracts and builds an RBAC taxonomy consisting of 49 role-permission pairs from the top 1,000 pairs mined.
\name then utilizes the mined common RBAC practices as a ``knowledge base for AC repair'' to guide LLMs in fixing code with similar functionality.
To help LLMs understand the functionality of the vulnerable code, \name employs static code slicing to extract AC-related code context, more specifically, an AC context graph (ACG).
With this two-fold source of information, \tool instructs GPT-4 to follow the Chain-of-Thought (CoT)~\cite{wei2022chain} prompting to identify the proper role-permission pairs. 
Eventually, \tool generates the patch and validates it according to the original vulnerability description.

We conducted evaluations comparing \name with SOTA tools~\cite{so2023smartfix, nguyen2021sguard} and performed an ablation study to highlight the improvements of individual components \name offers over the baseline GPT-4.
To comprehensively evaluate repair tools, we collected and constructed a benchmark dataset consisting of 118 cases from real-world attacks and contracts.
To the best of our knowledge, this is the first benchmark dataset specifically for AC vulnerabilities.
Our results showed that \name successfully repaired $94.92\%$ of AC vulnerabilities using appropriate AC mechanisms.
The ablation study further revealed that without the enriched context and mined taxonomy supplied by \name, vanilla GPT-4 fixed $52.54\%$ of vulnerabilities. 
\val agent further boosted the fixing rate from  $87.28\%$ to $94.92\%$.
Additionally, we analyzed the repair capabilities of tools across various role-permission pairs by category as well as their monetary and time costs.

Furthermore, to understand the value of \tool's repairs and how they differ from human repairs, we conducted a human-based evaluation involving 10 experts who have worked on smart contract auditing for 2-7 years.
The results show that \tool's repairs are mostly aligned with those of humans and are even finer-grained than those of both senior and junior experts, although in rare cases (3/118), human experts are better at handling open issues based on their knowledge and experience without much guidance.
Moreover, around half of the AC fixes are non-trivial to devise by humans, indicating that \tool can provide a unique complement to assist human-in-the-loop repair as a copilot.

\noindent
\textbf{Contributions.}
To sum up, our contributions are as follows:
\begin{itemize}[leftmargin=*, topsep=0pt, itemsep=0pt]
   \item We proposed \tool, the first tool designed to repair AC vulnerabilities by guiding LLMs to appropriately enforce RBAC mechanisms across a variety of scenarios.

   \item We assembled the first benchmark dataset of 118 AC vulnerabilities, sourced from real-world attacks and contracts, based on which, we conducted an extensive evaluation of the effectiveness and efficiency of \tool and SOTA tools and LLMs, including an ablation study.

   \item We obtained a taxonomy of common RBAC practices, including 49 role-permission pairs summarized from the top 1K unique samples mined from 344,251 on-chain contracts.

   \item We carried out a human study to understand the value of \tool's repairs, yielding new insights into the comparison between LLM-based and human repairs. 
\end{itemize}

%% file: rev_tex_2/2-motivation.tex
\begin{figure}[t!]
    \centering
    \begin{lstlisting}[language=Solidity, escapechar=\%]
function depositFromOtherContract(uint256 _depositAmount, uint8 _periodId, 
    bool isUnlocked, address _from
) external  { //vulnerable, fixed by onlyBank 
    require(isPoolActive,'Not running yet');
    _autoDeposit(_depositAmount,_periodId,isUnlocked,_from);
}
    \end{lstlisting}
    \vspace{-2ex}
    \caption{An example of smart contract AC vulnerabilities.}
    \label{lst:motivationexample}
\end{figure}

\begin{figure*}[!t]
    \centering
  \includegraphics[width=0.9\linewidth]{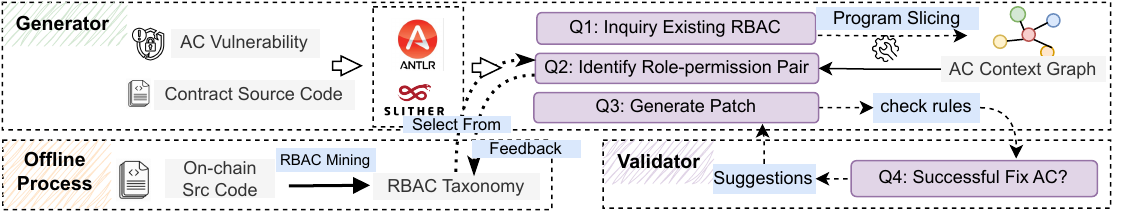}
    \centering
    \vspace{-2ex}
    \caption{A high-level overview of \tool, consisting of both offline and online phases.} 
  \label{fig:overview}
\end{figure*}
\smallskip

\section{Background and Motivation}
\label{sec:motivate}

\subsection{Background}
\noindent \textbf{Large Language Model.}
Pre-trained language models such as BERT~\cite{devlin2019bertpretrainingdeepbidirectional} and GPT~\cite{radford2018improving} have revolutionized the field of natural language processing (NLP) through pre-training on large text corpora.
This approach has enabled these models to develop robust, transferable language representations that are highly effective across a wide range of NLP applications.
Based on our evaluation of four popular LLMs, including GPT-4~\cite{chatgpt}, GPT-3.5~\cite{gpt35}, Mistral~\cite{jiang2023mistral7b}, and Llama3~\cite{llama3}, in \mysec\ref{sec:llm_selection}, we eventually use GPT-4 as the base mode..

\noindent \textbf{Smart Contract.} 
Smart contracts are self-executing agreements where the terms are encapsulated in executable code and run on blockchains~\cite{Szabo1997}.
However, smart contracts may be susceptible to software vulnerabilities, leading to financial risks.
If a contract allows for unauthorized ERC20~\cite{erc20} token transfers, a flaw like improper access control can expose it to risks such as malicious abuse of legitimate functions.

\noindent \textbf{Role-based Access Control} (RBAC)~\cite{sandhu1998role} is a well-known security paradigm in which \textit{permissions} are assigned to \textit{roles} rather than directly to users.
Each user belongs to one or more roles to accomplish various access control policies.
This approach encapsulates a set of permissions within each role, defining the actions a user can perform.
Nowadays, RBAC is recommended as the state-of-the-art security practice for separating the execution of access control policies from the management of business logic in smart contracts, usually through a set of well-defined modifiers~\cite{fang2023beyond, ozac}.

\subsection{A Motivating Example}
Our approach was motivated by a real-world AC attack on the DeFi application named \textit{GYMNetwork}\cite{defihackexample, example}.
\myfig~\ref{lst:motivationexample} shows the vulnerable function \texttt{depositFromOtherContract}, 
the root cause of which is that it is marked as \texttt{external}.
Without the validation by an appropriate modifier, an attacker was able to deposit numerous fake tokens to falsify his token shares in \textit{GYMNetwork}, leading to a loss of two million USD in 2022.

The patch provided by the original author added a modifier, \texttt{onlyBank}, to ensure that only the vault address can deposit tokens.
Since the role \texttt{Bank} had already been defined in the vulnerable contract, RBAC was partially implemented by the author previously.
In this case, the vulnerable function could have been repaired with existing RBAC mechanisms from the code context, by \texttt{onlyBank}, in accordance with the \textit{plastic surgery hypothesis}~\cite{xia2023revisiting}.
If the context is not considered during the repair, existing tools, such as SmartFix~\cite{so2023smartfix}, and LLMs (GPT-4) adopted conservative measures, i.e. \texttt{owner of the contract}, as in \S\ref{sec:RQ2}, which could lead to overfitting by inappropriately preventing legitimate banks from depositing.
Clearly, this not the expected behavior, as such repairs significantly impede the function's usability.
Instead, the appropriate repair should respect common RBAC practices and align with the context related to the access control of smart contracts.

\rev{
Similar to the motivating example, RBAC is commonly implemented in smart contracts through mechanisms such as centralized role mappings (e.g., \texttt{mapping(address => bool)}), modifier-based enforcement (e.g., \texttt{onlyOwner}), and inline conditional checks using \texttt{msg.sender}. These implementations often vary significantly across contracts in structure, naming conventions, and enforcement logic. This diversity introduces challenges for automated repair, including difficulty in identifying roles due to inconsistent definitions, implicit permission logic, and the risk of introducing conflicting or redundant access checks.
}

\subsection{Inspired Design of \tool}
\minor{
To address the heterogeneity of AC practices in smart contracts, \tool first mines common RBAC patterns from large-scale contracts and generalizes them into domain knowledge, organized as a dynamic taxonomy of role-permission pairs. The use of an external knowledge base to guide or supplement large language models is a widely recognized strategy for improving robustness and accuracy, as evidenced by recent advances in retrieval-augmented and knowledge-augmented LLMs~\cite{Schick2023Toolformer, Lewis2020RAG, baek2023knowledge}. Our RBAC taxonomy is designed to be continuously extensible, enabling the system to incorporate new knowledge as it encounters novel contexts. This taxonomy serves as a domain-specific external knowledge base, effectively grounding the LLM’s reasoning and mitigating risks of hallucination or inconsistency. By correlating the code context of an AC-related vulnerability with the taxonomy, the LLM can infer and apply AC mechanisms that are both contextually relevant and consistent with the contract’s intended logic.
}

\rev{
Importantly, \tool is also RBAC-aware—it analyzes the existing enforcement pattern within the contract and adapts its patching style accordingly. For instance, if a contract predominantly uses modifier-based enforcement, \tool will attempt to follow this style in the generated patch to maintain semantic consistency. This adaptive behavior helps prevent structural conflicts and promotes compatibility with the original design.
}

\rev{
Furthermore, the independent Validator Agent (\val) verifies whether the proposed patch aligns with the intended access control policy, ensuring that the role-permission relationship is correctly enforced based on RBAC and that the patch does not introduce functional regressions. This dual-layered approach—combining LLM-based reasoning guided by domain knowledge and semantic validation by \val—enables \tool to effectively address the challenges of RBAC heterogeneity in smart contract repair.
More details of this process, including the construction of the RBAC taxonomy, are further elaborated in Section~IV.
}

%% file: rev_tex_2/a-overview.tex
\section{Overview of \tool}
\label{sec:overview}

\myfig~\ref{fig:overview} presents a high-level overview of \name, which includes both offline and online phases.
In the offline phase, we mine common RBAC practices from smart contracts to construct an RBAC taxonomy.
This taxonomy will be used in the online phase to guide GPT-4 in pinpointing the appropriate role-permission pairs.
In the online phase, for each AC vulnerability, based on the Multi-Agent Debate (MAD) architecture\cite{liang2023encouraging, wang2023unleashing, du2023improving, xi2023rise}, we employ a dual-agent architecture that consists of a \gen and a \val.
Specifically, we mine the RBAC taxonomy from the source code of smart contracts deployed on-chain.
With this taxonomy in hand, \name repairs an AC vulnerability in the following steps:

\begin{enumerate}[leftmargin=15pt, topsep=0pt, itemsep=0pt]

    \item To facilitate practicality and avoid redundant fixes, an optional step involving a checking prompt Q0 is used to confirm if the target function is subject to AC vulnerabilities. This is because while \tool is positioned as an APR (Automatic Program Repair~\cite{goues2019automated}) tool that only takes confirmed vulnerability inputs from auditing reports, CVEs, and attack incidents, we allow \tool to be deployed as a copilot to help developers or existing AC detection tools fix potential AC vulnerabilities. In the latter case, Q0 is needed, and the result should be confirmed by an operator, as in the typical copilot scenario.

    \item \textit{Generator} then parses the contract source code, including the vulnerable part, to extract RBAC-related code elements.  We then provide these elements to GPT-4 in a prompt Q1, seeking to inquire whether any element belongs to existing RBAC mechanisms in the subject code.

    \item Starting from the vulnerable function $f_{vul}$, \gen employs program slicing and data flow analysis to construct an inter-procedural AC Context Graph (\rcg). This graph depicts the code semantically related to $f_{vul}$. Upon recognizing existing RBAC mechanisms in step (1), \gen extends the \rcg by incorporating relevant identifiers, such as modifiers and state variables, based on  $f_{vul}$.

    \item Using the serialized \rcg as prompt Q2, \gen guides LLMs to identify the most appropriate role-permission pair from our mined RBAC taxonomy or, if necessary, incorporates a new pair into the taxonomy.

    \item After pinpointing the role-permission pair, \gen instructs LLMs to generate a proper patch for the vulnerable code through prompt Q3. The generated patch is first statically checked for validity by rules and then continuously validated by \val through prompt Q4 to refine it until it is considered effective or the limit is reached.

\end{enumerate}

Next, we detail the offline phase of RBAC mining in \mysec\ref{sec:miningRBAC} and the online phase of RBAC-guided and context-aware LLM-driven repairing in \mysec\ref{sec:context} and \mysec\ref{sec:q3}, respectively. 
\rev{Regarding training or fine-tuning of LLMs, we observe that \tool already demonstrates robust performance in repairing AC vulnerabilities by leveraging rich contextual inputs and a comprehensive taxonomy. Although fine-tuning could potentially yield incremental improvements, it introduces risks of overfitting and may reduce model flexibility. In contrast, our current design enables \tool to dynamically incorporate newly identified RBAC pairs by updating the taxonomy, without necessitating retraining. Such adaptability cannot be achieved through fine-tuning a pre-trained model. Given these considerations, and the lack of large-scale training datasets for AC vulnerabilities, we designed \tool to effectively combine static analysis, a pre-defined taxonomy, and in-context learning prompts to repair AC vulnerabilities without additional training or fine-tuning.}

\rev{While ACFix leverages established techniques such as static slicing, code context extraction, and chain-of-thought prompting, its novelty lies in the domain-specific integration of these components to address the unique challenges of repairing AC vulnerabilities in smart contracts. Unlike general-purpose code repair, AC vulnerability repair demands precise reasoning over role-permission relationships and intricate identity checks. To address this, ACFix introduces a dynamic RBAC-guided taxonomy, a dual-agent validation-feedback mechanism, and a repair flow that directly consumes outputs from external AC detectors. This layered design enables ACFix to not only generate patches but also validate their semantic correctness and compatibility with existing RBAC logic, offering an end-to-end, practical solution tailored for secure smart contract repair.}

%% file: rev_tex_2/3-prestudy.tex
\section{Mining Common RBAC Practices}
\label{sec:miningRBAC}

\rev{During the offline phase, our goal is to systematically mine and categorize common RBAC practices observed in real-world smart contracts. Specifically, we extract role-permission pairs—the foundational elements of RBAC—from contract source code and generalize them into a structured taxonomy. This taxonomy serves as a domain-specific knowledge base that guides LLM-based repairs. By embedding this external knowledge, \tool can reason more effectively about access control logic and generate contextually accurate patches. Furthermore, the taxonomy is designed to be dynamically extensible at runtime, allowing \tool to incorporate new RBAC patterns as they emerge from usage examples.}

\begin{table}[t!]
\caption{A taxonomy of common RBAC practices, featuring mined role-permission pairs and their detailed checks.}
\label{tab:taxonomy}
    \vspace{-2ex}
\centering
\setlength{\tabcolsep}{0.6pt}
\resizebox{1\columnwidth}{!}{
\begin{tabular}{ccl}
\hline
\textbf{Roles} &
  \textbf{Permissions} &
  \multicolumn{1}{c}{\textbf{Examples of Detailed Permission Checks}} \\ \hline
 &
  \cellcolor[HTML]{EFEFEF}Low-level call &
  \cellcolor[HTML]{EFEFEF}Multi-factor authentication \\
 &
    \begin{tabular}[c]{@{}l@{}}Manage users \\ of the contract\end{tabular} &
  \begin{tabular}[c]{@{}l@{}}Multi-signature approval, Whitelisting and\\ blacklisting, Time locks\end{tabular} \\
 &
  \cellcolor[HTML]{EFEFEF}Manipulate price &
  \cellcolor[HTML]{EFEFEF}Rate limiting, Multi-signature requirements \\
 &
  Transaction management &
  Rate limiting, Transaction validation \\
 &
  \cellcolor[HTML]{EFEFEF}User/Role management &
  \cellcolor[HTML]{EFEFEF}Regular audits, Event logging for role changes \\
 &
  Utilities management &
  Time locks, Regular audits and testing \\
 &
  \cellcolor[HTML]{EFEFEF}Adjust fees &
  \cellcolor[HTML]{EFEFEF}Validation checks for fee changes \\
 &
   \begin{tabular}[c]{@{}c@{}}Monitor \& analyze \\ transactions\end{tabular} &
  \begin{tabular}[c]{@{}l@{}}Access control via view functions, \\Data validation and sanitation\end{tabular} \\
 &
  \cellcolor[HTML]{EFEFEF}Set trading pairs &
  \cellcolor[HTML]{EFEFEF}Validation checks for trading pairs \\
\multirow{-13}{*}{\textbf{Admin}} &
  Configure security settings &
  Multi-factor authentication \\ \hline
 &
  \cellcolor[HTML]{EFEFEF}Initialization &
  \cellcolor[HTML]{EFEFEF}\begin{tabular}[c]{@{}l@{}}Limit initialization to authorized users \\ against frontrun, Ensure initialization \\ only occurs once\end{tabular} \\
 &
  Change ownership &
  \begin{tabular}[c]{@{}l@{}}Limit ownership change to authorized users \\ against frontrun, Time locks\end{tabular} \\
 &
  \cellcolor[HTML]{EFEFEF}Upgrade contract &
  \cellcolor[HTML]{EFEFEF}\begin{tabular}[c]{@{}l@{}}Limit to authorized users against frontrun, \\ Time locks, Multi-signature requirements\end{tabular} \\
 &
  Pause contract &
  \begin{tabular}[c]{@{}l@{}}Limit to authorized users against frontrun, \\ Time locks\end{tabular} \\
  \multirow{-10}{*}{\textbf{\begin{tabular}[c]{@{}c@{}}Owner\\ of the \\ contract\end{tabular}}} &
  \cellcolor[HTML]{EFEFEF}Destroy contract &
  \cellcolor[HTML]{EFEFEF}\begin{tabular}[c]{@{}l@{}}Limit destroy to authorized users, \\ Multi-signature requirements\end{tabular} \\ \hline
 &
  Burn &
  \begin{tabular}[c]{@{}l@{}}Validation checks for the owner of \\ the burnable, Multi-signature control\end{tabular} \\
 &
  \cellcolor[HTML]{EFEFEF}Claim &
  \cellcolor[HTML]{EFEFEF}Validation checks for the owner of the claimable \\
 &
  Withdrawal &
  Rate limiting, Withdrawal limits \\
 &
  \cellcolor[HTML]{EFEFEF}Swap &
  \cellcolor[HTML]{EFEFEF}Transaction validation, Swap limits \\
 &
  Liquidify &
  \begin{tabular}[c]{@{}l@{}}Rate limiting, Validation checks for liquidified\\ funds\end{tabular} \\
 &
  \cellcolor[HTML]{EFEFEF}Transfer &
  \cellcolor[HTML]{EFEFEF}Validation checks for transferred funds \\
 &
  Approve &
  Validation checks for privilege of approver \\
 &
  \cellcolor[HTML]{EFEFEF}Manage stakes &
  \cellcolor[HTML]{EFEFEF}Validation checks for staking/unstaking \\
 &
  Create pools &
  Validation checks for pool creation \\
\multirow{-11}{*}{\textbf{\begin{tabular}[c]{@{}c@{}}Owner\\ of the\\ funds,\\ stakes, \\tokens\end{tabular}}} &
  \cellcolor[HTML]{EFEFEF}Set approval limits &
  \cellcolor[HTML]{EFEFEF}Rate limiting \\ \hline
 &
  Mint &
  \begin{tabular}[c]{@{}l@{}}Minting limits, Whitelisting and blacklisting, \\ Minter management, Multi-signature approval\end{tabular} \\
\multirow{-2}{*}{\textbf{Minter}} &
  \cellcolor[HTML]{EFEFEF}\begin{tabular}[c]{@{}c@{}}Setting minting\\ parameters \end{tabular}&
  \cellcolor[HTML]{EFEFEF}Validation checks for parameters \\ \hline
 &
  Offering loans &
  Validation checks for loan terms \\
 &
  \cellcolor[HTML]{EFEFEF}Collecting collateral &
  \cellcolor[HTML]{EFEFEF}Secure handling of collateral \\
 &
  Receiving payments &
  \begin{tabular}[c]{@{}l@{}}Transaction validation, \\ Secure mathematical operations\end{tabular} \\
 &
  \cellcolor[HTML]{EFEFEF}Managing defaults &
  \cellcolor[HTML]{EFEFEF}Secure collateral liquidation \\
 &
  Rolling loans &
  Validation checks for loan rollovers \\
 &
  \cellcolor[HTML]{EFEFEF}Withdrawal of funds &
  \cellcolor[HTML]{EFEFEF}\begin{tabular}[c]{@{}l@{}}Limit to fund owner, Withdrawal\\ limits, Time locks\end{tabular} \\
 &
  Viewing loan status &
  Data validation and sanitation \\
\multirow{-10}{*}{\textbf{Loaner}} &
  \cellcolor[HTML]{EFEFEF}Setting loan conditions &
  \cellcolor[HTML]{EFEFEF}Validation checks for loan conditions \\ \hline
 &
  Requesting loans &
  Validation checks for loan requests \\
 &
  \cellcolor[HTML]{EFEFEF}Depositing collateral &
  \cellcolor[HTML]{EFEFEF}Secure collateral handling \\
 &
  Repaying loans &
  Transaction validation, Secure math operations \\
 &
  \cellcolor[HTML]{EFEFEF}Managing active loans &
  \cellcolor[HTML]{EFEFEF}Data validation and sanitation \\
 &
  \begin{tabular}[c]{@{}c@{}}Rolling or \\refinancing loans\end{tabular} &
  Validation checks for rollovers/refinancing \\
 &
  \cellcolor[HTML]{EFEFEF}Handling liquidations &
  \cellcolor[HTML]{EFEFEF}Secure liquidation handling \\
 &
  Withdrawing collateral &
  Validation checks for withdrawals \\
\multirow{-9}{*}{\textbf{Borrower}} &
  \cellcolor[HTML]{EFEFEF}Receiving notifications &
  \cellcolor[HTML]{EFEFEF}Secure notification handling \\ \hline
 &
  Deposit &
  Restriction to owner of deposit, Deposit limits \\
 &
  \cellcolor[HTML]{EFEFEF}Withdrawal &
  \cellcolor[HTML]{EFEFEF}\begin{tabular}[c]{@{}l@{}}Withdrawal limits, Time locks, \\ Multi-signature approvals\end{tabular} \\
 &
  Manage funds &
  Rate limiting, Multi-signature approvals \\
\multirow{-4}{*}{\textbf{\begin{tabular}[c]{@{}c@{}}Vault,\\Bank\end{tabular}}} &
  \cellcolor[HTML]{EFEFEF}Set interest rates &
  \cellcolor[HTML]{EFEFEF}Validation checks for parameters \\ \hline
 &
  Log &
  \begin{tabular}[c]{@{}l@{}}Secure storage of sensitive information \\ Multi-signature requirements, Rate limiting\end{tabular} \\
\multirow{-2}{*}{\textbf{Logger}} &
  \cellcolor[HTML]{EFEFEF}Set log parameters &
  \cellcolor[HTML]{EFEFEF}\begin{tabular}[c]{@{}l@{}}Multi-signature requirements, Using proxy \\ patterns for upgradability and security\end{tabular} \\ \hline
\end{tabular}
}
\end{table}

To mine common RBAC practices, we have collected smart contracts written in Solidity~\cite{solidity} from $344,251$ addresses~\cite{tintinweb} on the Ethereum Mainnet as of December 2023.
While we found that developers often create their own versions of RBAC, there are three major mechanisms to enforce permission checks in smart contracts:

 \ding{172}\textbf{OZAC}: When OpenZeppelin Access Control (OZAC)~\cite{ozac} is employed, roles are explicitly and uniformly implemented using templates, such as \texttt{Ownable} and \texttt{Access}. We extracted the defined roles and corresponding function names based on OZAC templates to infer permissions.
 \ding{173}\textbf{Modifier}: Modifier declares conditional checks that Solidity automatically embeds into the function prologues~\cite{fang2023beyond}. However, since modifiers can be used for various purposes, we focused only on RBAC-related modifiers that begin with \texttt{only}, such as \texttt{onlyOwner}, based on an empirical study about modifiers~\cite{fang2023beyond}. The roles specified after \texttt{only} and the names of \textit{modified} functions were recognized as roles and permissions, respectively.
 \ding{174}\textbf{Transaction-Reverting Statements (TRS)}: The third is based on TRS~\cite{TRS21}, which use Solidity keywords, such as \texttt{require} and \texttt{if...revert}, to ensure contract integrity. A primary use of TRS is AC, where \texttt{msg.sender} is compared to predefined roles or addresses. Although TRS can serve multiple purposes, our study specifically targeted TRS assessing \texttt{msg.sender} in the context of RBAC, ensuring that our extraction remains relevant and omits distractions from unrelated uses of these statements.

\rev{Based on the three patterns above, we automatically mined $810,344$ pairs of roles and functions.
After de-duplication, we identified $46,495$ unique pairs, ranked in descending order by frequency.
To construct the RBAC taxonomy, we began by analyzing the top 1,000 most frequent role-permission pairs, which collectively account for 81.83\% of $810,344$ all observed pairs in our data. We employed an open card-sorting methodology~\cite{opencardsorting} to manually categorize permissions based on associated function names. New cards (i.e., role-permission categories) were dynamically introduced whenever a pair could not be reasonably grouped into an existing category.
}

\rev{
The first two authors, each with over four years of experience in smart contract analysis, independently reviewed and labeled all 1,000 pairs. After individual labeling, we first merged cards that conveyed the same underlying meaning. Then, we compared the assigned cards for each pair to identify disagreements. In cases of disagreement, the final decision was made by the third author. The overall disagreement rate was 9.5\%, indicating a high level of consistency between reviewers.
Following this process, every one of the 1,000 pairs was assigned to a specific role-permission card, and the resulting collection formed the foundation of our final RBAC taxonomy.
}

Table~\ref{tab:taxonomy} lists the categorized top mining results, with the first column showing the commonly used roles and the second column showing the permissions these roles may hold.
We notice that these role-permission pairs are mostly related to DeFi because AC is usually implemented to manage financial assets in smart contracts.
The roles could involve those with high privileges, such as \textit{Owner of the Contract} and \textit{Admin}, or those defined for specific operations, such as \textit{Minter} and \textit{Loaner}. The detailed roles depend on the usage of the contracts.
It is worth noting that initially, there were 48 role-permission pairs derived from on-chain contracts in the offline process.
Later during the evaluation, \tool dynamically updated the taxonomy and added one more pair, \textit{Admin-Low-level Call}.
The total of 49 pairs may not be exhaustive, but our evaluation showed that they have covered the majority of scenarios for which AC is implemented, and \tool could update it whenever new pairs are found (see Prompt Q2 in \mysec\ref{sec:q2}).

Based on the mined role-permission pairs, we further collected detailed permission checks for each pair from security auditing reports, as listed in the third column of Table~\ref{tab:taxonomy}, which provide examples of common RBAC practices.

\textbf{Revisiting the Motivating Example.}
With the derived taxonomy of common RBAC practices, we now revisit the motivating example in \myfig~\ref{lst:motivationexample} to intuitively demonstrate how this taxonomy could enable \tool to generate the appropriate roles and permissions for real-world vulnerable code.
Specifically, the function \texttt{depositFromOtherContract} could be easily matched by LLMs to the permission \texttt{Deposit} listed in Table~\ref{tab:taxonomy}.
Moreover, given the code context provided by our slicing in \mysec\ref{sec:context}, LLMs can determine that this vulnerable contract has implemented two RBAC role checks, \texttt{onlyBank} and \texttt{onlyOwner}.
Considering this context information and the taxonomy, LLMs could deduce the proper role-permission pair, which is \texttt{Bank-Deposit}, and generate a correct patch using the modifier \texttt{onlyBank} rather than \texttt{onlyOwner}.

%% file: rev_tex_2/4-methodology.tex
\section{Guiding LLMs to Pinpoint Proper Role-Permission Pairs Based on Code Context}
\label{sec:context}

With the common RBAC practices mined in \mysec\ref{sec:miningRBAC}, we now use them as a ``knowledge base for AC repair'' to guide LLMs in fixing code with similar functionality.
To help LLMs understand the functionality of subject vulnerable code that needs to be repaired, we employ static code slicing to extract AC-related code context, more specifically, an AC context graph (ACG).
We are particularly interested in code context related to the subject code's RBAC mechanisms.
Therefore, we first leverage LLMs to identify existing RBAC mechanisms in the subject code (\mysec\ref{sec:q1}), enrich the code context of the identified RBAC mechanisms into ACG (\mysec\ref{sec:ACG}), and finally instruct LLMs to use ACG to pinpoint the appropriate role-permission pair from the mined RBAC practices (\mysec\ref{sec:q2}).
During this process, we adopt the Chain-of-Thought (CoT)~\cite{wei2022chain} prompting to guide GPT-4 step by step, including the eventual AC repair generation that will be presented in the next section (\mysec\ref{sec:q3}).

\subsection{Identifying Existing RBAC Mechanisms}
\label{sec:q1}

To prevent conflicts with any pre-existing RBAC mechanisms and to guide the construction of a relevant ACG in subsequent steps, \tool employs GPT-4 to explore existing RBAC mechanisms in the subject code, given that GPT-4 can comprehend the code.
Since the names of most code elements, such as functions, state variables, and modifiers, are often self-explanatory, \tool extracts the names of these elements that might be associated with RBAC management.
This initial information, along with the source code of vulnerable function $f_{vul}$, is presented to GPT-4, which is then tasked with identifying the relevant elements related to RBAC.
Specifically, \tool first analyzes the contract to identify all pre-defined roles, permission checks, and enforcement mechanisms, including both modifier-based and inline conditional statements. \minor{If multiple AC mechanisms coexist within the same contract, \tool aggregates all detected enforcement styles and uses the combined RBAC structure as a reference for patch generation. When a new role-permission pair is inferred, \tool ensures it does not contradict or duplicate existing logic. If any overlap or redundancy is detected, the patch is generated to either update outdated logic or integrate seamlessly with the existing mechanisms in a non-redundant manner. The LLM is instructed to consider the complete set of enforcement styles to prevent the introduction of conflicting or inconsistent RBAC rules. This comprehensive analysis helps maintain a coherent and unified AC policy, even in scenarios involving multiple roles or complex, mixed enforcement implementations.}

\begin{figure*}[t!]
	\centering
	\includegraphics[width=0.9\linewidth]{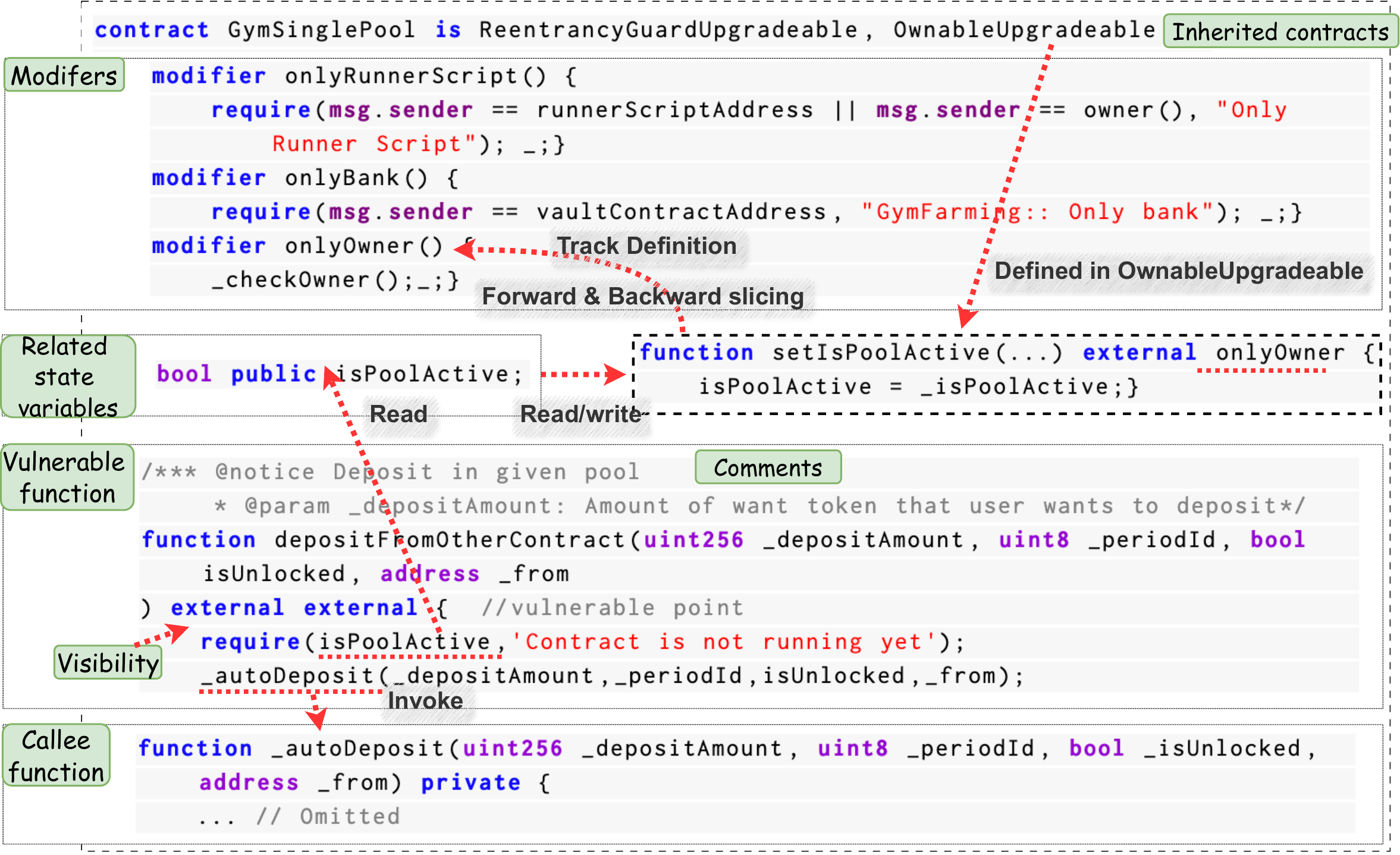}
	\caption{AC Context Graph (\rcg) for the Motivating Example.}
	\label{fig:rcg}
\end{figure*}

\begin{figure}[h]
  \includegraphics[width=0.99\linewidth]{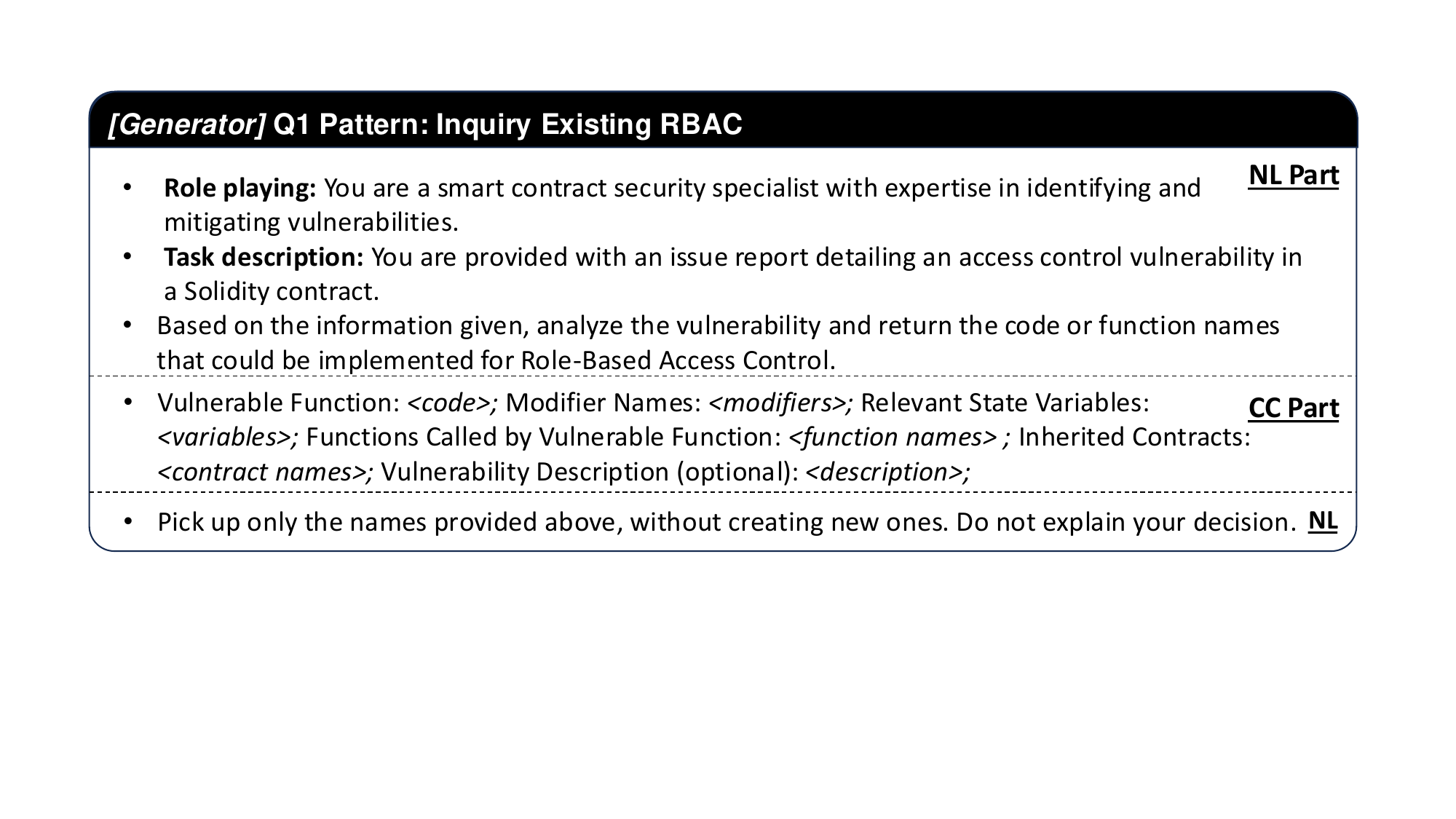}
  \caption{Q1 Prompt: Existing RBAC Identification}
  \label{fig:promptQ1}
\end{figure}

We designed our prompt based on the best practices commonly associated with using GPT-4, as suggested by~\cite{david2023you} and~\cite{LLM4Vuln24}.
Specifically, our prompt includes two parts: \ding{172} the natural language (\textbf{NL}) part that explains the task to GPT-4, and \ding{173} the code context (\textbf{CC}) part that contains the vulnerable function and other relevant code.
Given that the inquiry aims to identify RBAC-related code portions, \tool does not include detailed code statements but only the names of relevant functions and modifiers.
Following research on learning-based unit test generation~\cite{yuan2023no}, we include the following code context in the CC part:
(1) the signature and body of the vulnerable function; (2) modifiers; (3) state variables; (4) inherited contracts; (5) functions called by the vulnerable one in sequence; and (6) any vulnerability descriptions provided in the report, if available.
For the NL part, drawing upon widely recognized guidelines for using GPT-4~\cite{sun2023gpt, chen2023chatgpt}, we embed: (1) a role-playing instruction (i.e., \textit{You are a smart contract security specialist with expertise in identifying and mitigating
vulnerabilities}) to inspire GPT-4's contract repairing capability;
and (2) a task-description instruction 
to explain the task.
The prompt template is illustrated in Figure~\ref{fig:promptQ1} for Q1.

After pinpointing specific target elements, \tool constructs the \rcg based on them if available and $f_{vul}$ by default.

\subsection{Constructing AC Context Graph (ACG)}
\label{sec:ACG}

To capture contextual code statements that constitute the functionality of the vulnerable function $f_{vul}$, we employ program slicing~\cite{weiser1984program} as suggested by numerous previous studies~\cite{de1996understanding,badihi2020ardiff,badihi2023responsibility,xiao2020mvp,zhang2022has}.
Program slicing identifies code statements that influence, either through data or control, a target variable or statement.
Since Ethereum-compatible blockchains~\cite{BlockScope23} depend on modifications to state variables,
vulnerable functions generally interact with state variables in their own or other contracts, either directly or indirectly.
Based on this observation, \tool performs inter-procedural program slicing on the state variables interacted with by $f_{vul}$ and associated RBAC elements (i.e., the output of \mysec\ref{sec:q1}).
This approach aims to minimize extraneous code, ensuring a \textit{concise} prompt that attracts focused attention from GPT-4.
\tool, therefore, constructs an \rcg that comprises a streamlined code context of $f_{vul}$ from the subject contract.

\begin{algorithm2e}[th!]
	\small
	\SetAlgoNlRelativeSize{0}
	\SetNlSty{textbf}{(}{)}
	\SetCommentSty{mycommfont}
	
	\caption{Construction of AC Context Graph}
	\label{alg:1}
	\DontPrintSemicolon
	
	\KwIn{Vulnerable Function \(f_{\text{vul}}\), Program Dependency Graph \(\mathcal{PDG}\), Call Graph \(\mathcal{CG}\)}
	\KwOut{Access Control Context Graph \(\mathcal{G}=(\mathcal{V},\mathcal{E})\)}
	
	\(\mathcal{G} \gets \text{Graph}(\mathcal{V}, \mathcal{E})\) \tcp*[r]{Initialize Graph}
	\(\mathcal{V}_{\text{state}} \gets \text{DefUseChain}(f_{\text{vul}})\) \tcp*[r]{Extract State Variables}
	\(\mathcal{F}_{\text{callee}} \gets \text{CallGraph}(f_{\text{vul}})\) \tcp*[r]{Identify Callee Functions}
	\(\mathcal{V} \gets \mathcal{F}_{\text{callee}} \cup \mathcal{V}_{\text{state}}\) \tcp*[r]{Define Vertex Set}
	
	\ForEach{\(v \in \mathcal{V}_{\text{state}}\)}{
		\(\text{stmt} \gets \text{DefUseChain}(v)\) \tcp*[r]{Compute Def-Use Chain}
		\(f \gets \text{FuncOf}(\text{stmt})\) \tcp*[r]{Determine Enclosing Function}
		\(\mathcal{D}_{\text{data}}, \mathcal{D}_{\text{control}} \gets \{\text{stmt}\}, \{\text{stmt}\}\) \tcp*[r]{Initialize Dependencies}
		
		\While{\(\mathcal{D}_{\text{data}}.next() \neq null\)}{
			\(stmt \gets \mathcal{D}_{\text{data}}.next()\)\\
			\ForEach{\(\text{opr} \in stmt.split()\)}{
				\If{\(v \in \text{opr}\)}{
					\(f.\text{AddOperation}(\text{opr})\) \\
					\If{\(\mathcal{PDG}.\text{HasNextDataNode}(\text{opr})\)}{
						\(\mathcal{V}.\text{Add}(\mathcal{PDG}.\text{NextNode}(\text{opr}), f)\) 
						\(\mathcal{E}.\text{Add}(opr, \mathcal{PDG}.\text{NextNode}(\text{opr}))\) 
						\(\mathcal{D}_{\text{data}}.\text{Add}(\mathcal{PDG}.\text{NextNode}(\text{opr}))\) 
					}\ElseIf{\(\mathcal{PDG}.\text{NextDataNode}(\text{opr}) \in \{\text{PARAMETER}, \text{RETURN}\}\)}{
						\(\text{callsites} \gets \mathcal{CG}.\text{GetCallers}(\text{opr})\) 
						\(\mathcal{D}_{\text{data}}.\text{Add}(\text{Return}(\text{callsites}))\) 
					}
				}
			}
		}
		
		\While{\(\mathcal{D}_{\text{control}}.next() \neq null\)}{
			\(stmt \gets \mathcal{D}_{\text{control}}.next()\)\\
			\ForEach{\(\text{opr} \in stmt\)}{
				
				\If{\(v \in \text{opr}\)}{
					\(f.\text{AddOperation}(\text{opr})\)\\
					\If{\(\mathcal{PDG}.\text{HasNextControlNode}(\text{opr})\)}{
						\(\mathcal{V}.\text{Add}(\mathcal{PDG}.\text{NextNode}(\text{opr}), f)\) 
						\(\mathcal{E}.\text{Add}(opr, \mathcal{PDG}.\text{NextNode}(\text{opr}))\) 
						\(\mathcal{D}_{\text{control}}.\text{Add}(\mathcal{PDG}.\text{NextNode}(\text{opr}))\) 
					}\ElseIf{\(\mathcal{PDG}.\text{NextControlNode}(\text{opr}) \in \{\text{PARAMETER}, \text{RETURN}\}\)}{
						\(\text{callsites} \gets \mathcal{CG}.\text{GetCallers}(\text{opr})\) 
						\(\mathcal{D}_{\text{control}}.\text{Add}(\text{Return}(\text{callsites}))\) 
					}
				}
			}
		}
	}
	\KwRet{\(\mathcal{G}\)} \tcp*[r]{Return the constructed graph}
\end{algorithm2e}

We define \rcg as $ G = \{\left \langle V,E \right \rangle|V \subseteq \{F, Var_{state}, Mdf, Cmt\}, E \subseteq \{v_{i}, v_{j}\}| v_i, v_j \in \{f, var, mdf, cmt\}\}$, where $F$ represents the set of functions.
$Var_{state}$ denotes the set of state variables, $Mdf$ signifies the set of modifiers, and $Cmt$ is the set of comments.
Each vertex has three properties: \textit{Signature}, \textit{Body}, and the original \textit{Contract} to which it belongs.
Edges encapsulate multiple types of relationships between vertices, including \textit{invocation}, \textit{modifying}, \textit{reading/writing}, and \textit{comment}.
\myfig~\ref{fig:rcg} in Appendices presents an illustration of \rcg for the motivating example shown in \myfig~\ref{lst:motivationexample}.
Specifically, \tool breaks down the contract into various elements, such as modifiers and state variables, and connects them with corresponding relationships.
For individual processing of elements, \tool performs call-chain-based inter-procedural program slicing.

To facilitate the analysis, the call graph and Program Dependency Graph (PDG)~\cite{ferrante1987program} are firstly constructed.
Given that the input source code may not represent a complete Solidity project but rather excerpts from audit reports, it might not be compilable.
Hence, program analysis tools like Slither~\cite{feist2019slither} are not applicable due to their strict compilation requirements.
To address this issue, we have implemented a hybrid framework that performs call graph and PDG analysis on the Abstract Syntax Tree (AST) using Antlr~\cite{parr1995antlr} when Slither is infeasible.
Note that Intermediate Representation (IR) based analysis from Slither is preferred.
Although using Antlr may result in reduced accuracy and granularity (since AST primarily captures syntactic relationships between tokens without inherent optimization, unlike the IR-based approach), it remains adequate for collecting information for this task.

However, the usage of Antlr introduces two new issues.
First, unlike the three-address-code format in Slither IR, one-line source code format in Antlr might encompass multiple operators.
It is necessary to split multiple operations from one statement for proper slicing.
Second, it is common to accommodate the implementation within internal functions.

\rev{
To address these issues, we propose several enhancements for the construction of the \rcg. The general procedure of program slicing is presented in Algorithm~\ref{alg:1}. Initially, the Vulnerable Function $f_{vul}$, Program Dependency Graph $PDG$, and Call Graph $CG$ were first calculated based on the given contract serving as the basic structure to run the algorithm for inter-procedural construction. Specifically, the initial variables are extracted and initialized in Line 1-4 and \tool begins to iterate over the state variables $Var_{state}$ in Line 5. For each $var_{state}$, the statements that read or write the $var_{state}$ are tracked in Line 6 with the enclosing function being determined in Line 7. Next, \tool begins slicing from statements ($stmt$) involving the state variables $Var_{state}$ and conducts forward and backward slicing recursively by tracking dependencies related to these statements in the subsequent lines. If any operation is included in the slice, the corresponding complete line of source code is preserved in \textit{Body}.  
}

\rev{
During slicing, \tool recursively explores dependency chains using a Breadth-First Search (BFS) strategy, as illustrated in Lines 9–22. If a statement contains multiple operations (Line 11), it is split according to Solidity syntax using Antlr lexical patterns. Operations utilizing state variables are subsequently added to $f$ as initial points for data flow tracing (Lines 12–13). Then, \tool iteratively traces data and control flows, updating $D_{data}$ and $D_{control}$ accordingly (Lines 14–15 and 23–24).
For cross-function slicing, \tool connects the parameters at function call sites with their counterparts in the function definitions, enabling backward inter-procedural slicing (Lines 16–17 and 25–26). For forward slicing, the returned variable within the function definition is linked with variable assignments receiving the function's return value at call sites. For simplicity, Algorithm~\ref{alg:1} does not explicitly distinguish between forward and backward slicing.
}

\subsection{Pinpointing the Role-Permission Pair}
\label{sec:q2}

In this step, \tool leverages LLMs to correlate the enriched ACG code context with common RBAC practices to identify the role-permission pair for the subject code.
Due to the limited context window, \rcg is serialized as the prompt for GPT-4.
Specifically, elements from \rcg are described in both code segments and natural language and are presented to GPT-4.
\tool first supplements the source code body for modifiers.
For functions, only the statements derived from \rcg are included in the body code.
For state variables, the function bodies obtained from slicing are provided.
Regarding inherited contracts, such as \texttt{Ownable}, the bodies of modifiers defined therein are incorporated into the prompt.
In addition to these elements, edges, such as \textit{invocation}, \textit{modifying}, \textit{reading/writing}, and \textit{comment}, are all described in natural language.

\minor{Specifically, GPT-4 is prompted to select a role-permission pair from a pre-defined RBAC taxonomy. If GPT-4 identifies a pair that is not present in the current taxonomy but appears contextually appropriate, it is allowed to suggest a new pair. When such a novel pair is generated, \tool initiates a multi-stage validation process to ensure both its relevance and uniqueness. First, the system checks for potential duplication by comparing the candidate pair with existing entries using normalized role and permission representations. This normalization process standardizes role and permission names by converting them to a consistent case, removing common prefixes or suffixes, and applying stemming or lemmatization to address minor linguistic variations. In addition, synonyms and abbreviations are mapped to unified forms using a curated dictionary and context-aware LLM prompts. By leveraging these canonicalized representations, \tool can more accurately detect true semantic overlaps and avoid false positives. If no equivalent pair exists, the candidate is provisionally added to the taxonomy.
\\
\indent To further ensure the integrity and clarity of the RBAC taxonomy, \tool periodically employs an additional language model to systematically review the entire set of pairs. This review phase is designed to identify improper, overlapping, or ambiguous entries, and to sanitize the taxonomy if necessary. When appropriate, human oversight can be incorporated to prevent unintended errors and resolve borderline cases. This consolidated review process is conducted at regular intervals, balancing cost efficiency with the need for accuracy and minimizing disruption to ongoing repair operations.
\\
\indent For example, during evaluation, \tool encountered the pattern \textit{Admin–Low-level Call}, which was not present in the original taxonomy. Recognizing its contextual relevance, \tool successfully incorporated this pair into the taxonomy, making it available for subsequent repair tasks.
\\
\indent This dynamic extension, normalization, and validation mechanism enables \tool to adapt to diverse and evolving RBAC models across smart contracts, ensuring continued relevance and extensibility.}

Similar to the previous prompt, the prompt Q2 includes the CC and NL parts.
The CC part is detailed with \rcg information.
In the NL part, a question is posed to GPT-4, asking it to select a role-permission pair from the taxonomy based on the provided code context.
The prompt is as follows:

\begin{figure}[h]
  \includegraphics[width=0.99\linewidth]{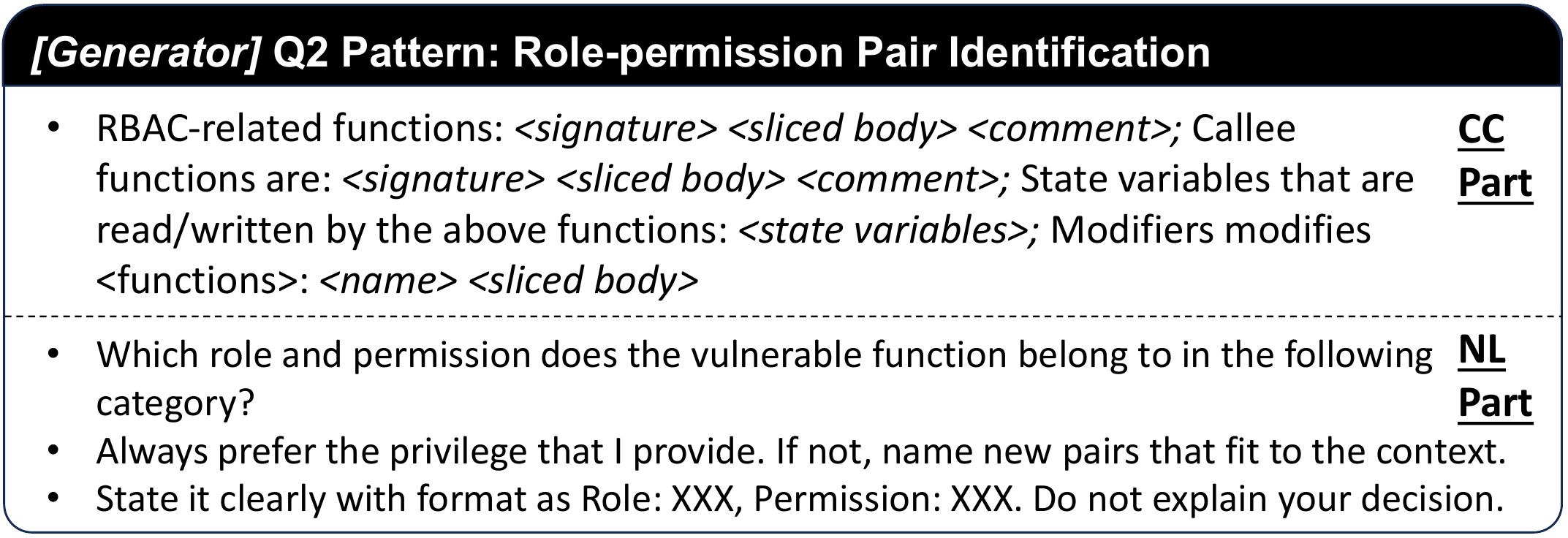}
  \caption{Q2 Prompt: Role-permission Pair Identification}
  \label{fig:promptQ2}
\end{figure}

%% file: rev_tex_2/b-patch.tex
\section{Generating and Validating Patches}
\label{sec:q3}

\subsection{Generating Patches and Static Checking}
\label{sec:verification}

With the appropriate role-permission pair identified in \mysec\ref{sec:context}, \name now generates the final AC repair.
Besides the role-permission pair stored in the LLMs' session memory from prompts Q1 and Q2, \name also retrieves corresponding examples of detailed permission checks from Table \ref{tab:taxonomy} to prompt GPT-4 to generate a patch.
If any existing RBAC mechanisms were identified in prior responses, \tool will prioritize reusing and enhancing them when possible to prevent any conflicts.
The prompt is presented as follows:

\begin{figure}[h]
  \includegraphics[width=0.99\linewidth]{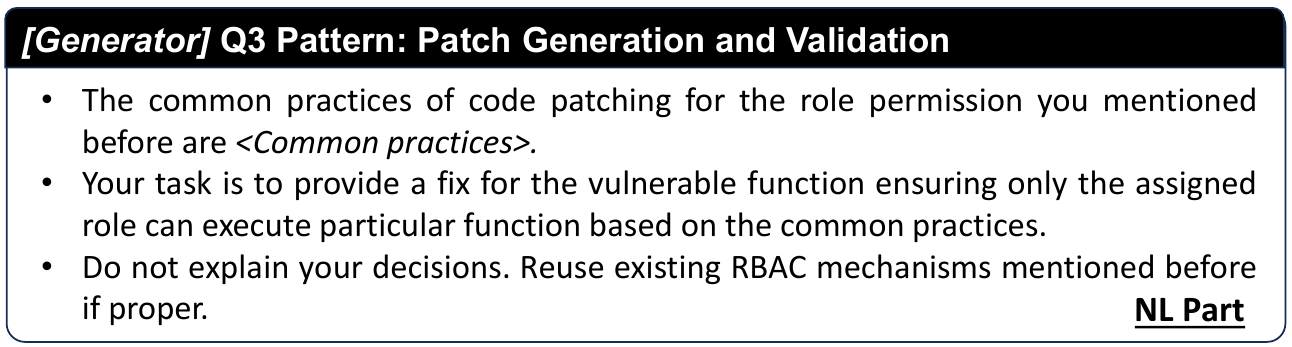}
  \caption{Q3 Prompt: Patch Generation and Validation}
  \label{fig:promptQ3}
\end{figure}

After deriving the repaired code, \tool conducts static grammar checks to ensure the validity of the repair.
Should any discrepancies arise, \tool consolidates these issues and relays them back to GPT-4 in a subsequent prompt, seeking an updated patch.
This paper considers five kinds of static grammar checks: Avoiding Undefined Tokens, Avoiding Infeasible Function Invocations, Avoiding Misused Types, Avoiding Inconsistent Solidity Versions, and Validating the \texttt{msg.sender} Check.
Details are omitted here due to page limit.
Interested readers may refer to our supplementary material.

\subsection{Generating Patches and Static Grammar Checking}
\rev{
After generating patches, \tool performs a series of static and semantic checks to ensure the compatibility, correctness, and applicability of the patches before integration into the target smart contract. These checks cover both syntactic and contextual dimensions, aiming to prevent invalid or incompatible modifications that could introduce unintended behaviors. The following rules are enforced:
\begin{itemize}[leftmargin=8pt]
	\item \textbf{Avoiding Undefined Tokens}: \tool first extracts all defined tokens from the current and inherited contracts, denoted as \(T_{\text{defined}}\). Then, it analyzes the tokens introduced in the generated patch, such as new functions, modifiers, and state variables, represented as \(T_{\text{repaired}}\). A patch passes this check only if all new tokens are properly defined or already exist.
	\begin{equation}
		\text{isDefined}(T_{\text{repaired}}) \Leftrightarrow T_{\text{repaired}} \subseteq (T_{\text{current}} \cup T_{\text{inherited}})
	\end{equation}
	\item \textbf{Avoiding Infeasible Function Invocations}: GPT-generated code may call functions that do not exist or have incorrect signatures. \tool collects the set of Solidity built-in functions \(F_{\text{built-in}}\) and the user-defined functions in the repaired contract \(F_{\text{repaired}}\), then validates that all invoked functions \(Invok\) are part of this union.
	\begin{equation}
		\text{isFeasible}(Invok) \Leftrightarrow Invok \subseteq (F_{\text{built-in}} \cup F_{\text{repaired}})
	\end{equation}
	\item \textbf{Avoiding Misused Types}: To prevent inconsistent or unsafe variable usage, \tool extracts the variable types from both the original (\(Type_{\text{vul}}\)) and repaired (\(Type_{\text{rep}}\)) contracts. It ensures that variable types are used consistently and that no invalid type conversions occur.
	\begin{equation}
		\text{isConsistent}(Type_{\text{vul}}, Type_{\text{rep}}) \Leftrightarrow Type_{\text{vul}} = Type_{\text{rep}}
	\end{equation}
	\item \textbf{Avoiding Inconsistent Solidity Versions}: A patch may use features unavailable in the specified version of Solidity. \tool checks whether the version required by the patch (\(Version_{\text{patch}}\)) is compatible with the contract's declared version (\(Version_{\text{sol}}\)).
	\begin{equation}
		SolCompa(Patch, SolVer) \Leftrightarrow Version_{\text{patch}} \subseteq Version_{\text{sol}}
	\end{equation}
	\item \textbf{Ensuring \texttt{msg.sender} Checks Are Introduced}: For access control enforcement, \tool verifies the existence of at least one conditional statement that compares \texttt{msg.sender} to a new or existing role identifier.
	\begin{equation}
		checked(msg.sender) \Leftrightarrow \exists\, if(msg.sender == role')
	\end{equation}
	\item \textbf{Def-Use Chain Validation}: \tool constructs def-use chains for all newly introduced or modified variables and ensures that each variable is correctly defined before use. This includes checking scope correctness, avoiding uninitialized variables, and ensuring no overwritten variables conflict with existing control/data flow.
        \begin{equation}
        \text{isValidDefUse}(V) \Leftrightarrow V_{\text{patch}}^{\text{use}} \subseteq (V_{\text{defined}} \cup V_{\text{patch}}^{\text{def}})
        \end{equation}
	\item \textbf{Structural Compatibility Check}: Before applying the patch, \tool validates that the modified code block aligns with the structural boundaries of the original smart contract. For example, a function-level patch must respect existing function signatures and modifiers.
        \begin{equation}
        \text{Compatible}(S_{\text{patch}}, S_{\text{target}}) \Leftrightarrow \forall s \in S_{\text{patch}}, \text{ValidContext}(s, S_{\text{target}}) 
        \end{equation}
\end{itemize}
This multi-step validation pipeline enables \tool to generate patches that are not only grammatically valid but also semantically consistent and directly applicable to the original smart contract codebase.
}

\subsection{Validating Patches' Effectiveness via MAD}
\label{sec:llmdebating}
\rev{
Once all static and rule-based checks are passed, \tool engages the Validator Agent (\val) to perform a higher-level semantic validation of the patch’s effectiveness through a multi-agent debating (MAD) loop. This step is essential to ensure not only syntactic correctness but also functional and security alignment with the intended AC policy.
In this process, the Generator Agent (\gen) first outputs a candidate patch and provides it to \val along with the vulnerability description, the surrounding code context, and the selected role-permission pair. The Validator Agent independently evaluates whether the patch (1) correctly mitigates the identified AC vulnerability, (2) preserves the original contract logic, and (3) does not introduce any new security or logical flaws.
}

\rev{
The \val performs this assessment by simulating the review process a domain expert might conduct. It reasons over the vulnerability description and the repaired code to determine if the AC logic is properly enforced—e.g., checking that access is restricted to intended roles, permission boundaries are respected, and the role-permission pair selected by \gen is consistent with the contextual semantics.
If the patch is deemed insufficient or flawed, \val returns structured feedback, including the reason for rejection (e.g., incorrect role, missing validation, logic conflict). This feedback is then passed to \gen, which uses the information to refine and regenerate an improved patch. This repair-validation cycle continues in a loop with a maximum of 3 iterations to balance thoroughness and efficiency.}

\rev{
Even if the patch is not accepted after 3 attempts, the last generated patch is retained as the final output. Based on our empirical evaluation (see Section~\ref{sec:RQ2}), this iterative mechanism proves highly effective: over 90.9\% of the cases required at most one re-attempt, and only a single case failed to pass validation after three rounds.
This agent-based validation framework strengthens \tool by introducing a self-regulating feedback loop that improves patch robustness, reduces hallucinations, and ensures more consistent adherence to access control principles.
}

\begin{figure}[h]
\includegraphics[width=0.99\linewidth]{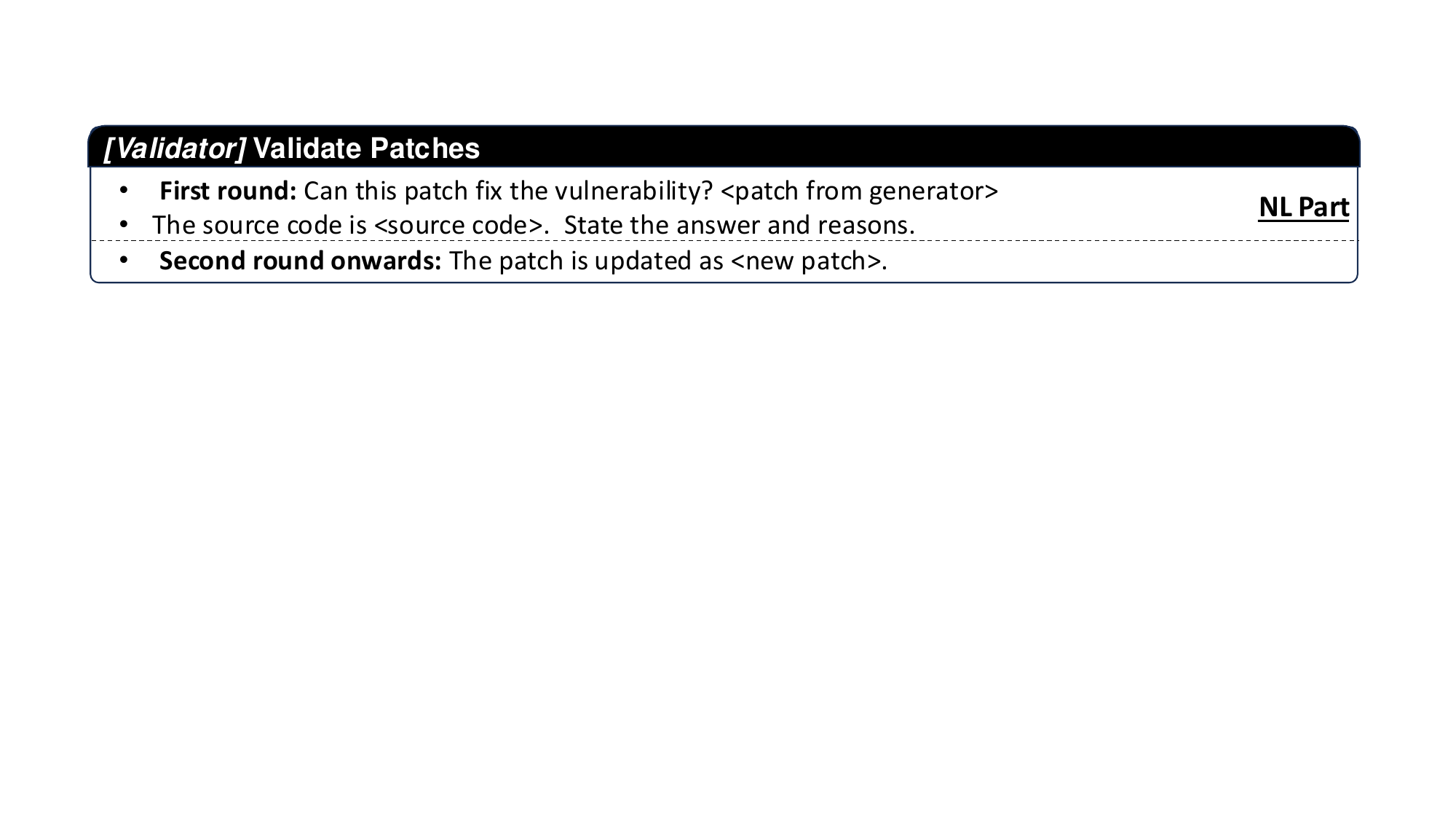}
  \caption{Q4 Prompt: Patch Validation}  
  \label{fig:prompt4}
\end{figure}

%% file: rev_tex_2/5-evaluation.tex
\section{Evaluation}
\label{sec:evaluate}

We aim to evaluate \name based on its effectiveness in appropriately repairing AC vulnerabilities by answering the following six research questions (RQs):

    \noindent \textbf{RQ1: LLM Selection.} \textit{How do popular LLMs perform as the base model of \tool and which is the best? }\\
    \rev{Given the emergence of multiple LLMs offering similar code generation capabilities, we first needed to evaluate these models to determine the most suitable base model for \tool. To achieve this, we assessed all popular LLMs available as of the submission date, comparing their performance as query interaction models within \tool using a benchmark dataset. Specifically, we focused on two primary metrics: generation rate and success rate.}
\\

	 \noindent \textbf{RQ2: Effectiveness Analysis.} \textit{How effectively does \tool repair AC vulnerabilities compare to other vulnerability repairing tools for smart contracts?}\\
     \rev{After selecting the best-performing model, we conducted a comprehensive and fair comparison with existing smart contract repair tools. To facilitate this, we first created the initial benchmark dataset specifically tailored for AC vulnerabilities. Using this dataset, we evaluated all available tools, identifying their strengths and weaknesses. Furthermore, we analyzed failure cases to understand and reveal the underlying reasons behind incorrect repairs.
     }

    \noindent \textbf{RQ3: Ablation Analysis.} \textit{How does the performance of \tool compared to a baseline that uses only GPT-4 with raw code and descriptions as input? }\\
    \rev{
    To evaluate the contribution of each procedure implemented in \tool, we systematically masked individual components to clearly highlight their respective impacts. Additionally, we compared \tool against the vanilla GPT-4 model to emphasize the effectiveness and the novel design beyond the capabilities of the base model.
    }

    \noindent  \textbf{RQ4: Effectiveness by Categories.} \textit{How do tools perform across various categories in the benchmark dataset?}\\
    \rev{
    The complexity of repairing AC vulnerabilities largely depends on accurately discerning nuanced differences between candidate roles and comprehending the contextual meaning within the code. Consequently, the difficulty of cases within our benchmark dataset varies significantly. To gain deeper insights into the performance of \tool across varying levels of difficulty, we further categorized and analyzed these cases based on their ground truth RBAC pairs.
    }

   \noindent \textbf{RQ5: Practicality Analysis.} \textit{Is \tool able to check potential vulnerabilities reported by static checkers?}\\
    \rev{
    Since \tool is designed to repair vulnerabilities reported by vulnerability detectors, we conducted this experiment specifically to evaluate such practical scenarios. To demonstrate its effectiveness, we integrated \tool with three widely-used static analysis tools capable of detecting AC vulnerabilities, thereby simulating an end-to-end workflow for vulnerability handling. This step evaluates the Q0 component of \tool, ensuring its practical capability to correctly process inputs provided by vulnerability detectors.
    }

    \noindent \textbf{RQ6: Efficiency Analysis.} \textit{How does \tool perform in terms of efficiency and financial cost?}\\
    \rev{Considering both execution time and monetary cost, an LLM-based tool such as \tool is expected to deliver efficient and cost-effective vulnerability repairs with notable results. Therefore, we conducted an end-to-end evaluation, systematically monitoring the execution time and monetary expenses associated with using \tool.}

\input{rev_tex_2/c-setup}

\noindent \textbf{Metrics.}
\label{sec:metrics}
Given that evaluating the correctness of patches remains a challenge in Automatic Program Repair~\cite{goues2019automated}, determining whether a repair is appropriate for the contract without overfitting involves leveraging multiple metrics to evaluate repairers.
The following metrics were used for evaluation:

\begin{itemize}[leftmargin=15pt, topsep=0pt, itemsep=0pt]
	\item \textbf{Comparison with Author Fixes}: Due to security concerns, many DeFi organizations and teams refrain from publishing the corrected code post-attack. We managed to collect $20$ real fixes by the original authors to serve as target repairs for these $20$ cases. Any patch that diverged from these original fixes was deemed unsuccessful.
 
    \item \textbf{Exploitation-Based Evaluation}: DeFi Hack Labs~\cite{defihacklabs} provides exploitation scripts that demonstrate how vulnerabilities can be exploited in a simulated environment, using authentic contracts sourced from the blockchain. We used these scripts to determine whether the vulnerability remains exploitable after the repair. We ran exploit scripts on both the original and repaired code to demonstrate that the repaired contracts are no longer exploitable. The logs for both of them are provided in our dataset~\cite{dataset}.
    
    \item \textbf{Manual Inspection}: The first two authors manually examined the repaired contracts to determine if the patch was appropriate. The third author made the final decision in the event of a disagreement. The explanatory notes are listed in our dataset~\cite{dataset}.
\end{itemize}

It is worth noting that our initial intention was to utilize detection tools to determine whether the AC vulnerability still existed after repairs.
However, no suitable tool was found to work properly for the cases within our dataset (except for 19 CVEs).
Specifically, AChecker~\cite{ghaleb2023achecker} works only for bytecode contracts.
When we ran AChecker against 43 compilable AC cases, only 3 were detected (with testing logs recorded on our website~\cite{dataset}), leading to its exclusion from the evaluation.
SPCon~\cite{liu2022finding} requires transaction history, and SoMo~\cite{fang2023beyond} targets only modifier-based AC vulnerabilities and has yet to release its source code.
As for other generic detectors such as Securify~\cite{securify_pub} and Slither~\cite{feist2019slither}, they require either compilable source code or precompiled bytecode, with the exception of SmartCheck~\cite{tikhomirov2018smartcheck}.
However, upon running SmartCheck on our dataset, we found that it generated many false alarms about other types of vulnerabilities but very few concerning AC, indicating its unsuitability for detecting AC vulnerabilities.

\noindent\textbf{SOTA Repair Tools to Be Evaluated.}
Various repair tools for smart contracts have been proposed in recent years.
We selected benchmark tools through a principled selection process.
Initially, we searched for papers using keywords such as ``smart contract'' and ``security'' in top-tier security/software engineering/programming language venues from the past three years (up to June 2024), yielding 268 papers on smart contract security.
Excluding papers unrelated to vulnerability fixing, 15 relevant papers were derived.

From these papers, we identified 9 baseline candidates, including SGuard~\cite{nguyen2021sguard}, SGuard+~\cite{gao2024sguard+}, SmartShield~\cite{zhang2020smartshield}, SCRepair~\cite{yu2020smart}, Elysium~\cite{ferreira2022elysium}, Aroc~\cite{Jin2022aroc}, HCC~\cite{giesen2022practical}, EVMPatch~\cite{rodler2021evmpatch}, SmartFix (2023)~\cite{so2023smartfix}, ContractTinker~\cite{wang2024contracttinker}, and LLMSmartSec~\cite{mothukuri2024llmsmartsec}.
We then excluded three tools that were either inapplicable for our patch generation or unavailable, and four tools that only work on bytecode, resulting in a final list of three repair tools for source code.
Specifically, the artifact for HCC is not available.
Since SmartShield, Aroc, and Elysium are designed exclusively for bytecode repair, they were omitted from our comparative study.
Meanwhile, SCRepair requires manually curated unit tests for patch generation, a resource that our dataset lacks.
Among these tools, only SGuard and SmartFix have available artifacts and are capable of accepting source code and repairing AC vulnerabilities, leading to their inclusion in our analysis.
\rev{ContractTinker is an LLM-based smart contract repair tool designed to handle various vulnerability types without being limited to a specific category. We included it in our evaluation by adapting its input pipeline to process vulnerability descriptions in our dataset. The modified ContractTinker is denoted as ContractTinker*.
LLMSmartSec is another recent LLM-based approach aimed at secure smart contract generation and repair. However, at the time of our evaluation, it lacked runnable artifacts and clear documentation, making reproduction infeasible. 
SGuard+ extends the rule-based repair engine of SGuard with enhanced capabilities. However, it does not provide public implementation or configuration files, making it impractical to replicate its repair logic without introducing bias. 
Therefore, among these, SmartFix, SGuard, and ContractTinker were included as baselines in our evaluation to ensure a fair and reproducible comparison.}

\input{rev_tex_2/d-LLMCompare}

\subsection{RQ2: Evaluating \tool and SOTA Tools}

\tool, SmartFix, and SGuard were run on our benchmark dataset to generate patches.
We first checked the compilability of the patches.
Then, we evaluated the correctness of the patches using three metrics.
We introduced the term \textit{Generate Rate} ($Rate_{gen}$) to denote the percentage of generated patches across all cases, and 
\textit{Success Rate} ($Rate_{success}$) to represent the proportion of patches that meet the three criteria of the stipulated metrics as successful repairs.
As illustrated in Table~\ref{tab:bench}, \tool was able to generate patches for all 118 cases, with $112$ of them considered successful repairs, resulting in a \textit{Success Rate} of $94.92\%$.
In contrast, SGuard could only generate patches for $6$ case, and SmartFix for $21$ cases.
The analysis of results and reasons behind the performance of all tools will be elaborated upon.

\noindent
\textbf{Analysis of Results from \tool.}
Out of the 112 successfully repaired cases, their compilability was checked against 43 cases that were already compilable before patching.
It turned out that all of them could be successfully compiled with the corresponding Solidity versions.
As for the $6$ unsuccessful repair cases, we categorized them into four reasons:
(4 cases) \textbf{Over-protection} (overfitting) : \tool returned repairs that could potentially hinder the routine usage of certain users. For example, \tool repaired a contract that allowed anyone to steal the collateral of loaners by adding an \texttt{onlyOwner} modifier, which restricted access from normal loaners who were supposed to be authorized to claim their own collaterals. One case was caused by insufficient context provided from the context extraction step, so GPT-4 could not recognize the correct permissions. The other three were caused by the strict \val that prefers conservative measures.
(1 case) \textbf{Different from Real Fixes}: For most cases with real fixes, \tool performed well by providing the same protection as the real fixes. However, there was a case where the real fixes considered non-code information, which \tool could not predict from merely a code-based context. For example, the function \texttt{safeTransferFrom} was changed to \texttt{internal} from \texttt{external} after fixing, without any clear reason provided in the code.
    This change could potentially overfit against legitimate users.
(1 case) \textbf{Unclear Requirements of the Description}: The description of this vulnerability indicated only insufficient checks for potential users. Indeed, it required multiple checks for the arguments in addition to the \texttt{msg.sender} to ensure proper functionality. \tool failed to provide sufficient checks for this vulnerability.

\noindent
\textbf{Analysis of Results from SOTA Tools.}
As illustrated in Table~\ref{tab:bench}, SGuard~\cite{nguyen2021sguard} could only generate fixes for $6$ cases, and $1$ of them passed the three metrics.
SmartFix~\cite{so2023smartfix} managed to generate $21$ fixes with $7$ successful ones.
The primary reason for the failed cases of both tools is compilation failure because they depend on IR derived from compiled code.
However, sources for some AC vulnerability cases have not released on-chain addresses but only vulnerable code snippets.
Even when addresses are provided, the source code may not be disclosed by blockchain explorers such as Etherscan~\cite{etherscan}.
The analysis of the tools is elaborated as follows:

\noindent \textbf{SGuard}: All cases that were not generated were due to unsuccessful compilation, as logged by SGuard. Out of the 6 patches generated by SGuard, 5 failed to repair the AC vulnerability. Four of these failed cases had patches that were exactly the same as the vulnerable code, indicating that SGuard failed to identify the necessary fixes. For the remaining case, SGuard provided a fix that was irrelevant to AC. The only case correctly repaired involved the misuse of \texttt{tx.origin}, suggesting that SGuard was specifically designed to address \texttt{tx.origin} misuse in the context of AC vulnerabilities.
    
\begin{figure*}[!t]
	\centering
        
	\subfloat[\footnotesize \tool]{\includegraphics[width=.24\linewidth]{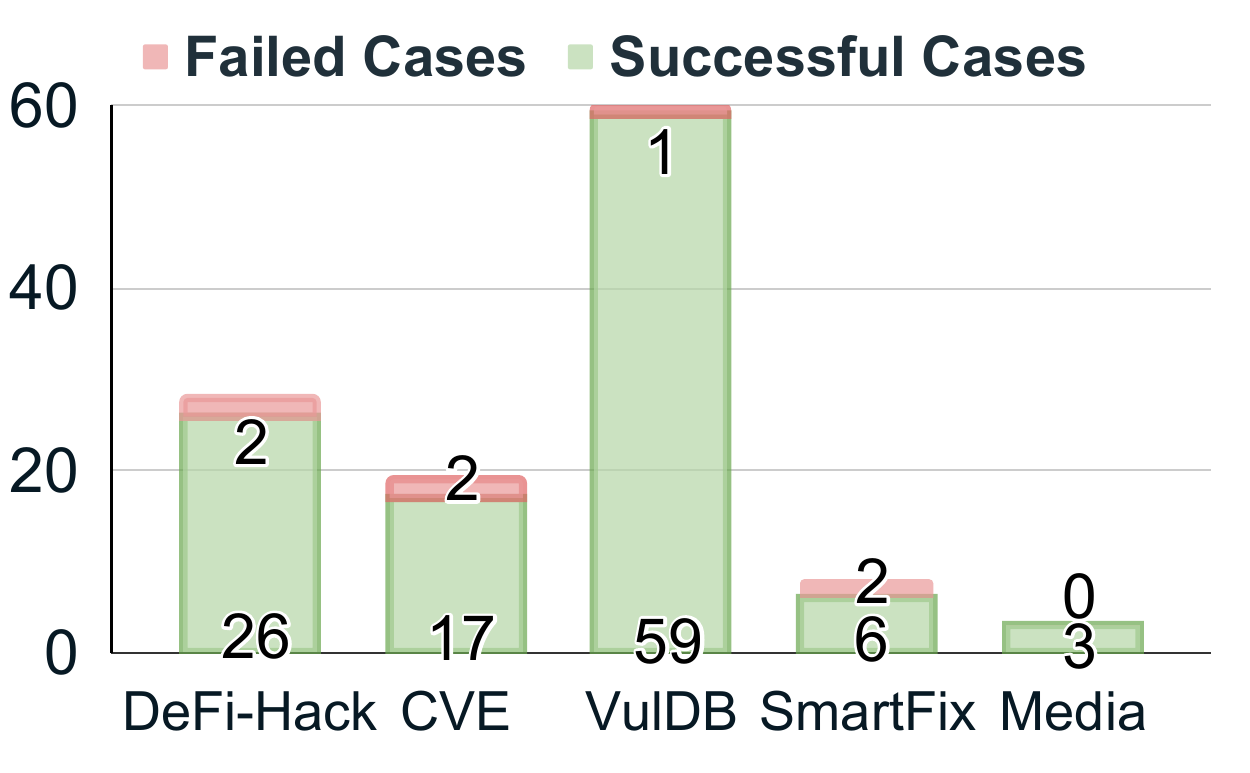}\label{fig:tool}}\hspace{1pt}
	\subfloat[\footnotesize \bla]{\includegraphics[width=.24\linewidth]{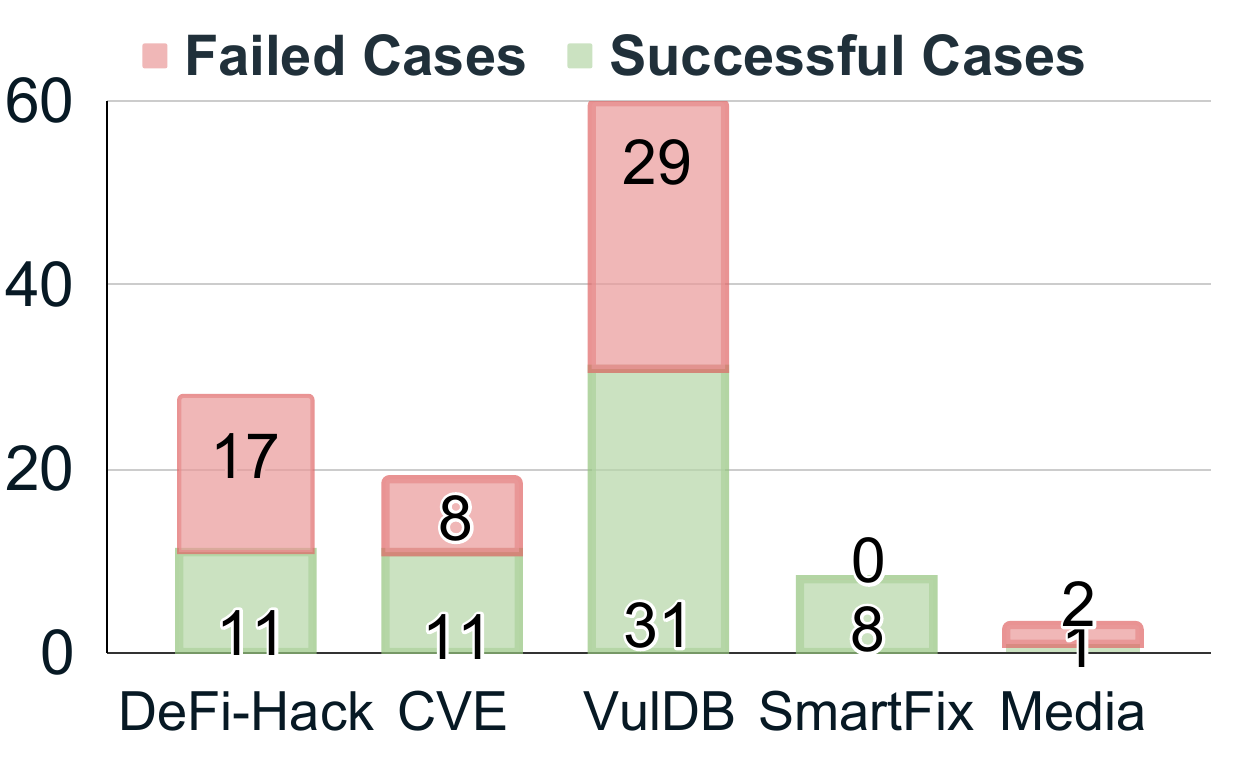}\label{fig:baseline}}\hspace{1pt}
	\subfloat[\footnotesize SGuard]{\includegraphics[width=.24\linewidth]{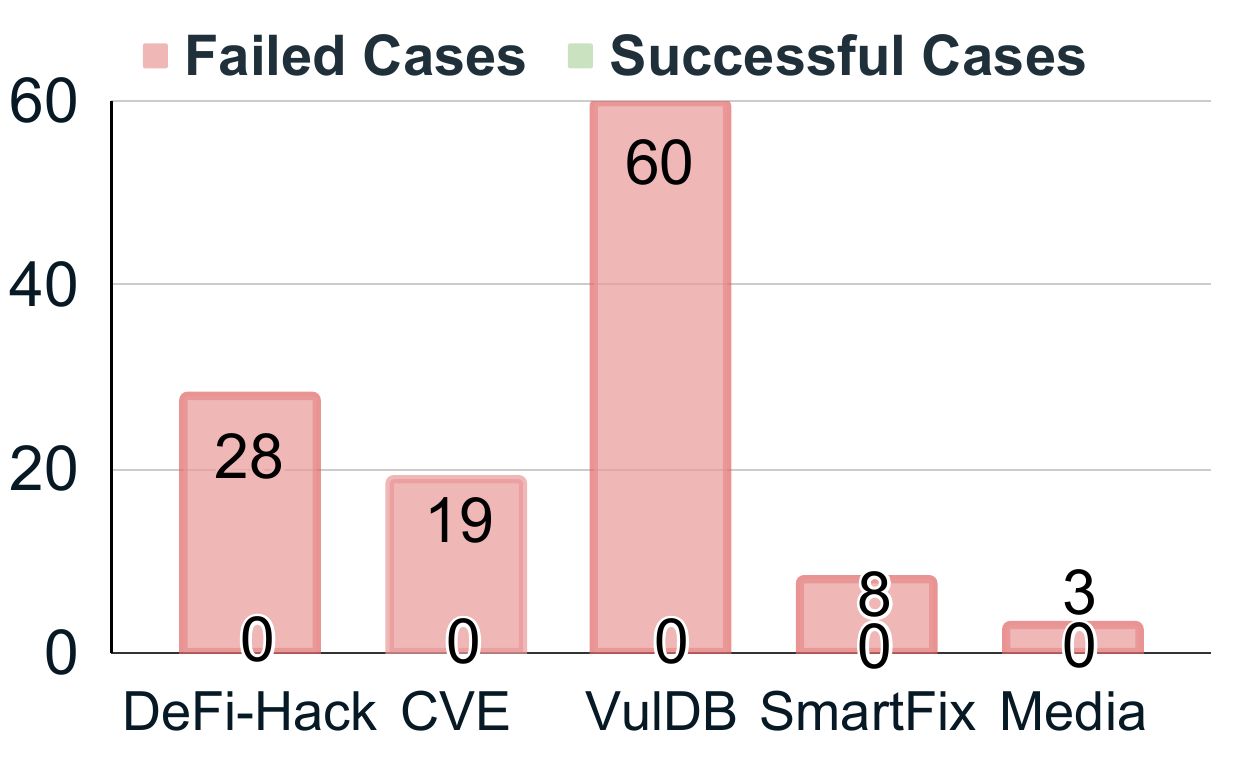}\label{fig:sguard}}
    \subfloat[\footnotesize SmartFix]{\includegraphics[width=.24\linewidth]{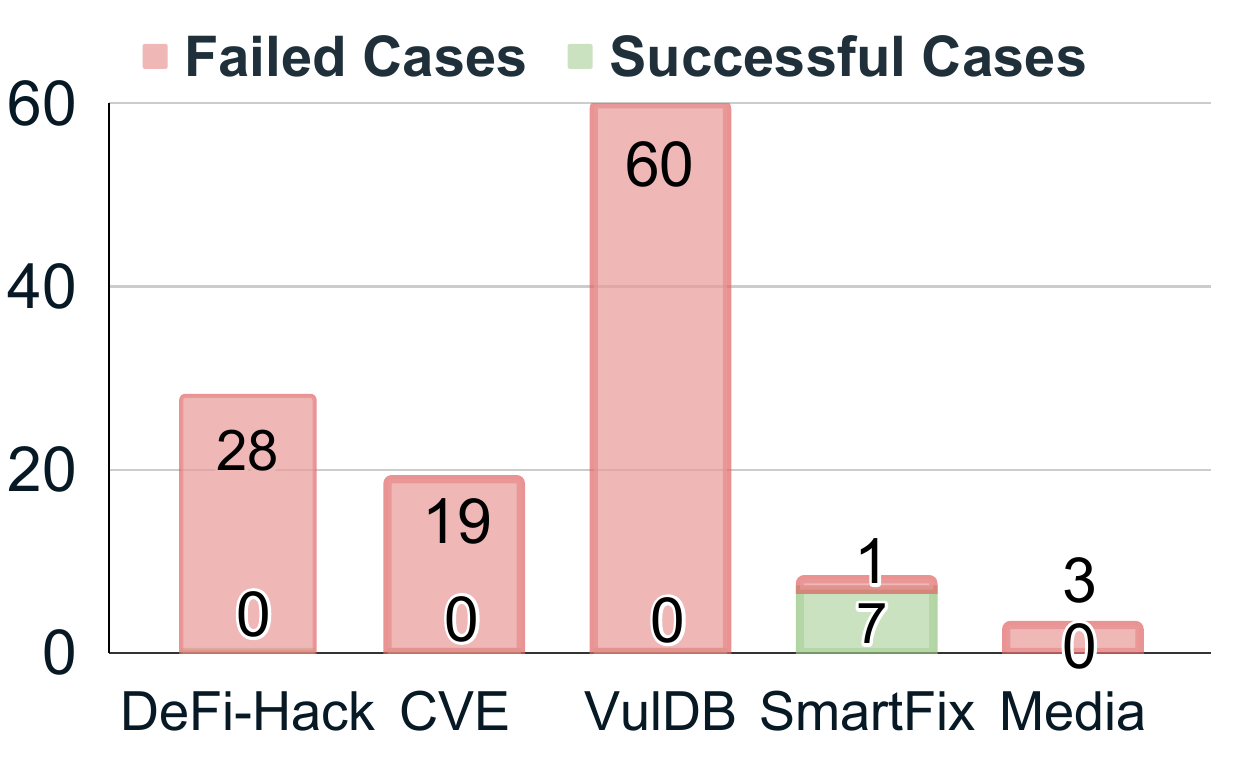}\label{fig:smartfix}}\hspace{1pt}\\
    \vspace{-2ex}
	\caption{Effectiveness of Tools on Various Data Sources.}
    \label{fig:effectiveness}
\end{figure*}

\noindent \textbf{SmartFix}: SmartFix generated patches for 21 cases, accounting for $17.80\%$ of AC vulnerabilities. However, only 7 of them successfully fixed AC vulnerabilities, all of which were cases of misuse of \texttt{tx.origin}. Among the unsuccessful repairs, none of the $14$ cases were related to \texttt{tx.origin} but to other types, as illustrated in \myfig~\ref{fig:smartfix}. Out of 14 unsuccessful patches, 13 targeted other non-AC vulnerabilities, including 12 cases of Integer Over/underflow and 1 case of Reentrancy, but left AC vulnerabilities unrepaired, which did not exist in the original contracts upon manually examination. SmartFix only accurately identified the AC vulnerabilities in two cases, both related to re-initialization issues. In these cases, SmartFix replaced the incorrectly named constructor function with the Solidity keyword \texttt{constructor}, without considering that the \texttt{pragma} versions were both \texttt{\string^4.x.x}, which does not support the \texttt{constructor} keyword. As this fix would lead to compilation failure, we labeled them as unsuccessful fixes. The overall result shows that SmartFix was designed to repair AC vulnerabilities, but its effectiveness is limited to types of AC such as re-initialization and misuse of \texttt{tx.origin}.

\noindent\rev{\textbf{ContractTinker*:}
Among the 118 cases in our dataset, only 43 contracts were compilable and thus eligible for processing due to ContractTinker*'s reliance on Slither for code analysis. Moreover, the tool expects structured vulnerability reports (e.g., `HighRiskFindings`), which were not consistently available in our dataset. After modifying its input pipeline to accept natural-language vulnerability descriptions, ContractTinker* generated patches for only 6 functions across 8 output files, accounting for just 5.08\% of the total dataset. Manual inspection of the outputs revealed that none of the generated patches correctly repaired the access control vulnerabilities. Only one fix is related to AC but produces an overfitting change, i.e., replacing `public` with `private`. The rest of the fixes were unrelated, including zero-address checks and balance checks. Therefore, the effective success rate of ContractTinker* in repairing AC vulnerabilities in our benchmark was 0\%. 
We attribute this low effectiveness to ContractTinker*'s design, which is built around a direct conversation with the LLM, without external knowledge integration or domain-specific context. As a result, its performance heavily depends on the availability of clean, structured vulnerability reports. However, in practice, such structured reports are often inconsistent or missing entirely, limiting the tool’s applicability in real-world repair scenarios.}

\begin{table}[]

\caption{Repair Results of Tools in the Benchmark Dataset.}
    \vspace{-2ex}
\centering
\resizebox{1\columnwidth}{!}{
\begin{tabular}{@{}lrrrr@{}}
\hline
\toprule
\textbf{Tool}                    & $\# Generated$ & $\#Success$ & $Rate_{gen}$       & $Rate_{success}$  \\ \midrule
\textbf{\tool}            & \textbf{118}     & \textbf{112}        & \textbf{100.00\%} & \textbf{94.92\%} \\ 
SGuard                     & 6               & 1                  & 5.08\%         & 0.85\%           \\ 
SmartFix                  & 21               & 7                 & 17.80\%             & 5.93\%               \\ 
\rev{ContractTinker*}                  & \rev{6}               & \rev{0}                 & \rev{5.08\%}             & \rev{0.00\%}               \\ 
\midrule
\blc                  & 113               & 106                & $95.76\%$                  & $89.83\%$          \\
\bld                  & 118               & 81                & $100.00\%$                  & $68.64\%$          \\
\blb                  & 118               & 103                & $100.00\%$                  & $87.28\%$          \\
\bla                  & 113               & 62                & $95.76\%$                   & $52.54\%$          \\

\bottomrule
\end{tabular}
}
\label{tab:bench}
\end{table}

\begin{tcolorbox}[boxrule=0.5pt,arc=1pt,boxsep=-1mm,breakable]
\textbf{Answer to RQ2}: \tool successfully generated repairs for $100\%$ of AC vulnerabilities, effectively fixing $112$ cases, representing a $94.92\%$ success rate. This demonstrates that \name can repair the majority of AC vulnerabilities across a variety of scenarios. It also outperforms SOTA contract repair tools, SGuard and SmartFix, which only successfully repaired the misuse of \texttt{tx.origin} and could not handle AC vulnerabilities in broader scenarios.
\end{tcolorbox}

\subsection{RQ3: Ablation Study}
\label{sec:RQ2}

\begin{figure}[h]
	\includegraphics[width=0.99\linewidth]{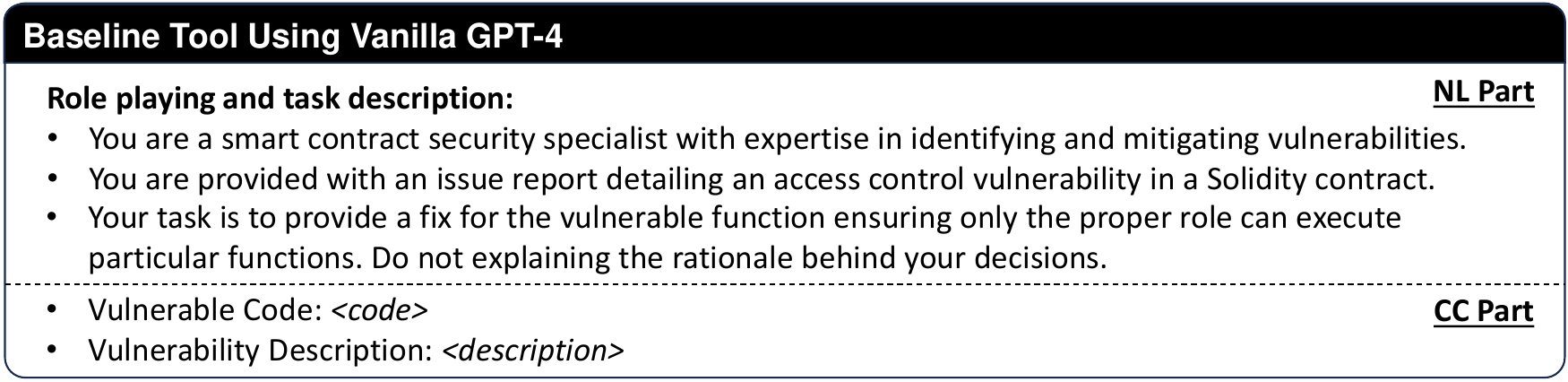}
	 \caption{Baseline Tool Using Vanilla GPT-4}
	\label{fig:promptBaseline}
\end{figure}

To demonstrate the effectiveness of the RBAC taxonomy, context information, and the MAD mechanism, we conducted an ablation study based on four customized baselines.
We iteratively removed individual components of \tool. Specifically,
\blc has the same implementation as \tool but without \rcg.
\bld lacks the RBAC taxonomy.
\blb solely utilizes \gen without \val.
\bla uses \texttt{GPT-4-0613} with raw vulnerable code and vulnerability descriptions directly, without preprocessing, as in Figure~\ref{fig:promptBaseline}.

As shown in Table~\ref{tab:bench}, \blc and \bla generated patches for 113 cases instead of 118 because five contracts have multiple source code files that exceeded the token limit.
Furthermore, \blc fixed 106 cases, indicating that \rcg contributed to 12 more successful cases.
The number is not significant as most of the contracts were retrieved from auditing reports, which have limited length.
However, \tool could benefit more from \rcg in terms of scalability for real-world deployed contracts.
The 81 successfully fixed cases demonstrate that RBAC taxonomy has significantly contributed to the patch generation. The taxonomy can be dynamically updated when new pairs are encountered by \tool, such that \textit{Admin-Low-level Call} was added by \tool during evaluation. 
It has been substantiated that GPT performs better for patch generation if the generation is guided by well-structured knowledge.

\begin{tcolorbox}[boxrule=0.5pt,arc=1pt,boxsep=-1mm,breakable]
\textbf{Answer to RQ3 for \blc and \bld}: The comparison with these two baselines have substantiated that \rcg and the RBAC taxonomy could improve the repair by fixing an additional 12 and 37 cases, respectively.
\end{tcolorbox}

For \blb, without \val, $15$ patches were not generated correctly.
It was observed that \val successfully validated $9$ more patches, resulting in correct patches.
The errors in 5 of these patches were previously due to misalignment with the vulnerability description, while another 4 were due to overfitting roles.
Fortunately, they were corrected after review by \val, meaning that MAD can effectively correct improper patches through independent evaluation.

Regarding the number of MAD loops,Within the 118 cases, \tool completed the generation after 0, 1, 2, and 3 re-attempts for 41, 68, 7, and 2 cases, respectively.
$92.37\%$ of cases were completed within 1 attempt.
This demonstrates that MAD typically converges quickly within 3 loops.

However, \val was observed to introduce over-fitting patches in some instances, in addition to correcting others.
\tool failed in $3$ cases due to over-fitting checks.
After scrutinizing the history of debates between agents, it was found that the patches were initially correct as generated by \gen.
However, \gen was persuaded to adopt conservative roles like \texttt{owner} by \val after debate.
Therefore, even with \val, determining the appropriate role-permission pair is stillchallenging.
Still, \val could effectively safeguard the output according to the evaluation.

\begin{tcolorbox}[boxrule=0.5pt,arc=1pt,boxsep=-1mm,breakable]
\textbf{Answer to RQ3 for \blb}: \blb failed to fix 9 cases compared to \tool, suggesting that $Rate_{success}$ could be further boosted with  \val.
\end{tcolorbox}

\bla has successfully repaired 62 cases ($52.54\%$).
We manually analyzed the distribution of the repaired cases and found that \bla tends to apply conservative roles in the repairs ($68.64\%$ of the total), such as \texttt{onlyOwner}.
For 40 out of the $60$ successful cases, \bla generated repairs using \texttt{onlyOwner}.
In another 17 cases, the ideal roles were specified in the vulnerability descriptions, allowing \bla to directly reuse the given roles.
For the remaining 3 cases, the function signatures themselves provided enough context for GPT-4 to infer potential roles, such as \textit{borrower} from the function \texttt{borrow}.
In contrast, out of the $58$ incorrect repairs, $37$ were inaccurately over-protected by \texttt{onlyOwner}, affecting legitimate users.
The rest of the cases were deemed improper because they were either still vulnerable or uncompilable. 

\begin{tcolorbox}[boxrule=0.5pt,arc=1pt,boxsep=-1mm,breakable]
\textbf{Answer to RQ3 for \bla}: Without the RBAC taxonomy and \rcg, \bla achieved a repair success rate of only $52.54\%$. This highlights the vital importance of the \rcg mined by \tool from the code and the guidance provided by the RBAC taxonomy.
\end{tcolorbox}

Besides the two baselines, we further explored the effectiveness of \gen rule checks.
Patches of $4$ cases violated the rules in \mysec\ref{sec:verification}.
Upon manual inspection, it was determined that 2 cases involved incompatible pragma versions, and the other 2 were related to mis-spelled variable names, which could be attributed to LLM hallucination or loss of focus~\cite{tian2023chatgpt}.
However, they did not affect the effectiveness of \tool, considering that the rule checks could safeguard the output.

\subsection{RQ4: Effectiveness by Categories of Roles}

\begin{figure}[ht]
  \vspace{-3ex}
  \includegraphics[width=1\linewidth]{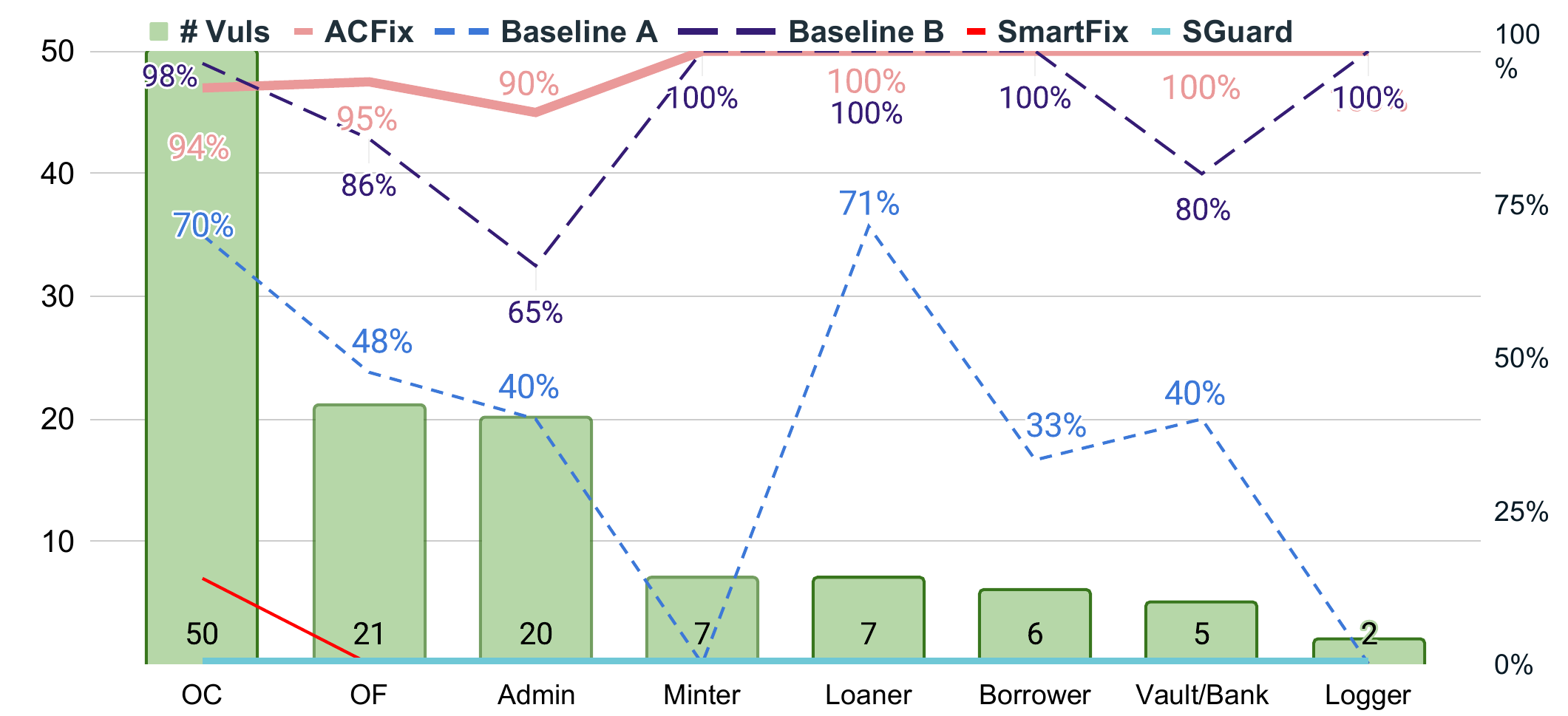}
  \vspace{-4ex}
  \caption{Proportion of successful repairs by roles.}
  \label{fig:rsum}
\end{figure}

\rev{The complexity of repairing AC vulnerabilities depends largely on accurately distinguishing nuanced role differences and interpreting the code context, resulting in varied difficulty levels across our benchmark dataset. To better understand \tool's performance under different complexity scenarios, we categorized and analyzed benchmark cases based on the roles.
After manually annotating the appropriate role-permission pairs for each vulnerability in the benchmark dataset, we further categorized the 118 AC vulnerabilities according to their corresponding roles.}

\rev{In this evaluation, the ground truths of our benchmark dataset were mapped to 31 out of 49 entries of the taxonomy, demonstrating that our taxonomy captures a wide range of AC patterns and is not overfitted to a narrow scope. This further confirms the dataset’s diversity and the generality of the taxonomy itself.}
As permissions may vary from case to case, we focused \myfig~\ref{fig:rsum} solely on eight major roles regarding the proportion of successful repairs. 
It was observed that three major roles—\textit{Owner of the Contract} (OC), \textit{Owner of Funds} (OF), and \textit{Admin}—account for the majority ($77.12\%$) of the AC vulnerability benchmark dataset.
Generally, \tool achieved the best repairs across the eight roles, but its performance for the roles of \textit{OC} and \textit{Admin} was less effective.
These roles usually have the broadest range of permissions, and \val tends to encourage \gen to adopt conservative roles, such as \textit{OC} and \textit{Admin}.
This is evidenced by \blb, which achieved slightly better results for the role of \textit{OC} (98\% v.s. 94\%).
Since \bla lacks refined context, it performs worse than \tool across all roles.

\begin{tcolorbox}[boxrule=0.5pt,arc=1pt,boxsep=-1mm,breakable]
\textbf{Answer to RQ4}:
\tool struggled with the roles of \textit{OC} and \textit{Admin} but still outperformed \bla across all roles.
On the other hand, SmartFix was only able to repair $17\%$ of the \textit{Initialization} cases. 
\end{tcolorbox}

\subsection{RQ5: Practicality Analysis}
\rev{
As described in \mysec\ref{sec:overview}, \tool is designed to function as a copilot for processing vulnerability reports generated by external AC vulnerability detectors. It is not intended to serve as a standalone detector, instead, it operates downstream by consuming flagged outputs, such as potentially vulnerable functions, and assisting in precise vulnerability localization and automated patch generation. Given the high computational cost of large-scale contract analysis, \tool is deliberately scoped to handle a limited set of contracts identified by existing detection tools. To evaluate its effectiveness in this role, particularly its ability to accurately confirm AC vulnerabilities and reduce human verification effort, we assessed the Q0 step of \tool on cases reported by three state-of-the-art AC vulnerability detectors.
}

\rev{
We selected three tools that publicly provide labeled TP and FP cases: Slither~\cite{feist2019slither}, SoMo~\cite{fang2023beyond}, and SPCon~\cite{liu2022finding}. AChecker~\cite{ghaleb2023achecker} was excluded due to the lack of a publicly available dataset. For true positives, we carefully curated 21 TP cases confirmed by the original tools in their released datasets. Since these tools independently verified these vulnerabilities, they serve as a reliable basis for TP evaluation.
}

\rev{
To simulate integration with vulnerability scanners, we provided \tool with only the vulnerable functions (as detected by the tools) and no further textual description. \tool was then tasked with identifying the vulnerable lines and determining whether the root cause was a missing or incorrect identity check (e.g., \texttt{msg.sender}). Successful identification under this setting validates \tool’s capability to localize and explain the vulnerability, making it a practical companion to detection tools.
}

\rev{
Regarding false positives, due to the scarcity of public FP cases from SoMo, we collected 10 cases from its dataset. To maintain balance, we randomly selected 10 FP cases each from Slither and SPCon, resulting in a total of 30 FP cases. Combined with the 21 TPs, the full evaluation consists of 51 unique cases. This expanded and clarified evaluation setup strengthens the reliability and interpretability of RQ5.
}

\begin{table}[!t]
\centering
\small
\caption{The Q0 analysis results for 40 positive (TP/FP) cases reported by three SOTA AC vulnerability detectors.}
    \vspace{-2ex}
\setlength{\tabcolsep}{4pt}
\begin{tabular}{lllll}
\hline
\toprule
\textbf{Detectors}     & \textbf{All TP} & \textbf{Slither FP} & \textbf{SoMo FP} & \textbf{SPCon FP} \\ \midrule
Correct/Total & 21/21  & 7/10       & 10/10   & 9/10    \\
\bottomrule
\end{tabular}
\label{tab:fp}
\end{table}

As shown in Table~\ref{tab:fp}, the Q0 step of \tool has been evaluated on these 40 positive cases.
If Q0 is able to accurately determine the actual vulnerability status of each case, the case is considered correctly identified by \tool.
Specifically, the case is correct if Q0 returns True for a TP case and False for an FP case.
It is shown that \tool could correctly identify most cases (36/40).
Although 4 FP cases were not correctly identified by Q0, no TP cases were missed by \tool, thus ensuring that real AC vulnerabilities could be fixed.

\begin{tcolorbox}[boxrule=0.5pt,arc=1pt,boxsep=-1mm,breakable]
\textbf{Answer to RQ5}: Out of 40 positive AC vulnerabilities reported by SOTA tools, \tool correctly identified 36 of them, demonstrating its practical value in confirming and fixing AC vulnerabilities.
\end{tcolorbox}

\subsection{RQ6: Cost Efficiency and Performance}

Table~\ref{tab:cost} shows the monetary and time costs of using \tool for all AC cases in the dataset.
Regarding monetary cost,
the average number of tokens used across two agents by \tool is $1,956.35$.
According to the current pricing plan~\cite{oldGPT4cost}, the average cost for repairing one AC vulnerability is $0.0587$ USD.
Consequently, repairing all vulnerabilities in the dataset costs a total of $6.9266$ USD.
Note that the token counts for \tool and \blb were much higher than for \bla because the costs of failed cases were not counted, and consecutive conversations require incorporating the previous history into the new prompt, which results in repeated counting of tokens.
Additionally, it is evident that \blc consumed more tokens than \tool because the raw source code was not processed to highlight critical information by constructing \rcg.
Instead, the raw source code of contracts was incorporated in the prompt, including redundant tokens, leading to unnecessary costs and inefficiency.
Regarding temporal cost, the average time for \tool to patch each AC case is 30.58 seconds.
The time required for static analysis may vary depending on each case's complexity.

\begin{tcolorbox}[boxrule=0.5pt,arc=1pt,boxsep=-1mm,breakable]
\textbf{Answer to RQ6}: On average, \tool costs 0.06 USD and takes 30.58 seconds per case.
\end{tcolorbox}

\begin{table}[!t]
\caption{Monetary and Temporal Costs of \tool.} 
    \vspace{-2ex}
\setlength{\tabcolsep}{1pt}
\resizebox{1\columnwidth}{!}{
\begin{tabular}{@{}lccccc@{}}
\hline
\toprule
\textbf{Name} & \textbf{Avg. Token}             & \textbf{Avg. Total Price (USD)}       & \textbf{Avg. Total Time (s)}   \\ \midrule
\tool & 1,956.35          & 0.0588/6.9429  & 30.58/3,608.23     \\ 
\blc & 2,813.68           & 0.0843/9.9474  & 37.23/4,393.16\\
\bld & 1,845.84           & 0.0552/6.5341  & 29.35/3,463.31\\
\blb & 1,777.90           & 0.0546/6.4428  & 25.23/2,977.14\\
\bla & 378.79            & 0.0192/2.2542  & 7.66/903.98 \\

\bottomrule
\end{tabular}
}
\label{tab:cost}
\end{table}

%% file: rev_tex_2/c-setup.tex
\noindent
\textbf{Data Preparation.}

\rev{We constructed our dataset by building upon existing research, starting with 19 Common Vulnerability Enumerations (CVEs) that are frequently cited in prior AC-related studies~\cite{liu2022finding,ghaleb2023achecker,fang2023beyond}. For reference, SPCon and SoMo~\cite{liu2022finding,fang2023beyond} each evaluated on 44 cases, AChecker~\cite{ghaleb2023achecker} used 21 cases, and SmartFix~\cite{so2023smartfix} focused on just 12 AC-related cases. While the SmartBugs dataset~\cite{durieux2020empirical} is commonly used in broader smart contract vulnerability research, it was excluded from our evaluation due to the absence of ground truth annotations identifying whether the cases involve access control vulnerabilities. As our focus is specifically on AC vulnerabilities, such omissions make SmartBugs unsuitable for inclusion.}

However, relying solely on CVEs does not yield a comprehensive evaluation.
Given the absence of a benchmark dataset for AC vulnerabilities, we introduce the first benchmark dataset of real-world instances with ground truths.
\minor{This dataset has been assembled from five primary sources as indicated in Table~\ref{tab:datasource}} (already covering the sources from the above-mentioned work):
\ding{172} 19 CVEs from NVD~\cite{nvd}. \ding{173} Defi Hack Labs~\cite{defihacklabs} has published numerous vulnerabilities with real-world attacks.
Under the ``Access Control'' category, we collected 28 cases with vulnerable code snippets and blockchain addresses.
\ding{174} An open vulnerability dataset provided by tintinweb~\cite{tintinwebvul} contains $28,699$ vulnerabilities sourced from real-world auditing reports.
After filtering for ``Access Control,'' we identified 60 unique cases.
\ding{175} The dataset from SmartFix~\cite{so2023smartfix} includes 8 AC cases related to the misuse of \texttt{tx.origin}.
\ding{176} Additionally, we collected $3$ more cases from media sources, including BlockSec~\cite{blocksec}, SlowMist~\cite{slowmist}, and Medium~\cite{medium}.
In total, we have compiled $118$ real-world cases, making it the most extensive publicly available AC vulnerability dataset to date~\cite{dataset}. 

\begin{table}[]

\caption{Sources of the Benchmark Dataset}
\setlength{\tabcolsep}{5pt}
\small
\begin{tabular}{@{}l|rrrrr@{}}
\toprule
{Source} & {NVD} & {DefiHackLabs} & {tintinweb} & {SmartFix} & {Media} \\ \midrule
{Count}  & {19~(\cite{nvd})}  & {28~(\cite{defihacklabs})}             & {60~(\cite{tintinwebvul})}                 & {8~(\cite{so2023smartfix})}        & {3}     \\ \bottomrule
\end{tabular}
\label{tab:datasource}
\end{table}

%% file: rev_tex_2/d-LLMCompare.tex
\subsection{RQ1: Pilot Study to Identify Suitable LLM}
\label{sec:llm_selection}

\begin{table}[!t]
\centering
\small
\setlength{\tabcolsep}{4pt}
\caption{Repair Results for the Popular Base LLMs.}
    \vspace{-2ex}
\begin{tabular}{@{}lrrrrccccc@{}}
\hline
\toprule

\textbf{Model}          & $\#Generated$  & $\#Success$    & $Rate_{gen}$  & $Rate_{success}$ \\ \midrule
\textbf{GPT-4} & \textbf{118} & \textbf{112} & \textbf{100.00\%} & \textbf{94.92\%}  \\
GPT-3.5        & 115          & 66           & 97.46\%        & 55.93\%           \\
Mistral-7b     & 113          & 58           & 95.76\%        & 49.15\%           \\
Llama3-8b      & 117          & 87           & 99.15\%        & 73.72\%     \\
\rev{Llama3.2-11b}      & \rev{118}          & \rev{95}           & \rev{100.00\%}        & \rev{80.51\%}     \\
\bottomrule
\end{tabular}
\label{tab:llm}
\begin{tablenotes}
\item \footnotesize{$\#Generated$ is the number of cases in which patches were generated. $\#Success$ is the number of cases that a correct patch is successfully generated passing 3 metrics.}

\end{tablenotes}
\end{table}

\rev{
Given the various available LLMs, we first tested several popular and state-of-the-art models, including GPT-4~\cite{chatgpt}, GPT-3.5~\cite{gpt35}, Mistral~\cite{jiang2023mistral7b}, and LLaMA 3~\cite{llama3}, to select the base LLM for \tool. In our updated evaluation, we further integrated LLaMA 3.2–11B, the latest lightweight variant of the LLaMA family, to enable a fair and up-to-date comparison with GPT-4.
All LLMs were implemented under the same evaluation pipeline, with the only differences being in output formatting. The OpenAI API allows for structured response formatting~\cite{openaiFormatting}, making output parsing straightforward for GPT-4 and GPT-3.5. In contrast, although we explicitly instructed Mistral and LLaMA models to respond in JSON format, consistent compliance could not be guaranteed. Therefore, we implemented a robust string-based parser to reliably extract structured outputs across all models.
We chose not to include GPT-4o in this comparison due to potential concerns around training-time data leakage and limited control over evaluation consistency. Including such models may lead to non-reproducible or unfair results, particularly in security-sensitive tasks like access control repair. The comparative results between GPT-4 and LLaMA 3.2–11B are highlighted in Table~\ref{tab:llm}, showcasing their respective performance in terms of patch accuracy, runtime, and model responsiveness.
}

To avoid data leakage, we selected the LLMs with the earliest cutoff dates.
For GPT-4, the model was \texttt{GPT-4-0613} (training data up to September 2021).
GPT-3.5 was \texttt{GPT-3.5-turbo} (also up to September 2021).
As Mistral and Llama3 were released more recently, the earliest models that we could find were from October 2023 and May 2024, respectively.
Therefore, these two models were trained with newer data, potentially leading to data leakage and enhancing their capabilities in evaluation.
The configurations for these LLMs were all set to a temperature of 0 (to suppress randomness) and a maximum of 4096 tokens for output.

Table~\ref{tab:llm} presents the results of comparison among LLMs.
The GPT-4 model has demonstrated an excellent ability to provide correct patches for AC vulnerabilities, as evidenced by its much higher $Rate_{success}$.
This proficiency stems from its reasoning ability to deduce the proper role-permission pairs.
After manually reviewing the failed cases of other LLMs, most were found to be caused by over-fitting roles, such as \textit{the owner of the contract}.
The difference in selected role-permission pairs among LLMs has exhibited their varying abilities to summarize roles by understanding the source code context.
Another major category of failed cases resulted from grammar mistakes leading to uncompilability.
Cases where patches were not generated were caused by requests rejected by the LLMs to generate patches.
Based on the overall results and the above analysis, other models except GPT-4 exhibit sub-optimal context comprehending, unreliable generated code, and rejected requests. Thus we selected GPT-4 as the base model for \tool, as mentioned earlier in \mysec\ref{sec:motivate}.

\begin{tcolorbox}[boxrule=0.5pt,arc=1pt,boxsep=-1mm,breakable]
\textbf{Answer to RQ1}: Given its superior performance in generating appropriate patches for AC vulnerabilities compared to three other popular LLMs, GPT-4 was chosen as the base model for \name.
\end{tcolorbox}

%% file: rev_tex_2/6-humanstudy.tex
\section{LLM-based Repair vs. Human Repair}
\label{sec:human}

Following the evaluation of \tool itself in \mysec\ref{sec:evaluate}, we further proceed to understand the value of \name's repairs from a human perspective, e.g., how they align with human repairs and whether they are non-trivial to devise by humans (if non-trivial, this means that \tool provides a unique complement to assist human-in-the-loop repair as a copilot).

Towards this objective, we conducted a human-based evaluation involving 10 practioners who have worked on smart contract auditing for 2-7 years.
They are divided into junior (2-4 years) and senior (4-7 years) groups.
Given the raw source code and vulnerability description, the participants were asked two questions:
\ding{172} Write down the most appropriate role-permission pairs they thought fit the situation;
\ding{173} Indicate if the patch is straightforward to come up with based on their understanding.
For \ding{173}, unless explicitly stated, they were not asked to produce patches but only to assess the difficulty, because manually curated patches are hard to normalize for comparison.
Note that as this study does not involve any personally identifiable information, the IRB (Institutional Review Board) requirement was waived by our institution.

\subsection{How \name's Repairs Align with Humans}
\label{sec:align}

After manually scrutinizing the role-permission pairs curated by experts, 83 (74\%) and 69 (62\%) pairs by senior and junior experts respectively were aligned with pairs produced by \tool for 112 corrected cases.
Despite the different proportions, we carefully reviewed the curated pairs and derived several findings.
\ding{172} Humans are more likely to reuse existing roles if the provided source code is not lengthy.
Hence, the pairs are mostly aligned for cases with existing roles defined in the source code.
\ding{173} Humans tend to reuse function names as permissions without distilling them into abstract permissions as in our RBAC taxonomy.
This phenomenon is especially evident for junior experts.
It might indicate that humans require training and experience to accurately identify and summarize the proper role-permission pairs for correct patches.
\ding{174} Humans are inclined to give conservative roles, such as \texttt{owner}, \texttt{admin}, and \textit{authorized user}.
On average, 96.4 (81.69\%) and 79.0 (66.95\%) of the roles given by junior and senior experts respectively were conservative, contrasting with the 64 (54.24\%) returned by \tool.

These experts were further asked to draft patches for 6 failed cases by \tool.
After validating their patches with the same metrics, it turned out that half of their patches were correct according to all metrics.
We took \textit{Quixotic}~\cite{quitoxic} as an example to demonstrate the difference. 
Its brief vulnerability description is \textit{Quixotic checks only the buyer's signature.}
The vulnerable function \texttt{fillSellOrder} has multiple arguments and only checks the buyer's identity. 
Human experts were able to construct patches involving the AC checks against \texttt{buyer} as well as other necessary argument checks, such as \texttt{expiration} and \texttt{price}, according to their auditing experience.
However, \tool strictly included only the \texttt{buyer} role without flexibly involving other necessary checks.

\begin{tcolorbox}[boxrule=0.5pt,arc=1pt,boxsep=-1mm,breakable]
\textbf{Takeaway}:
    \tool's repairs are mostly aligned with those of humans and are finer-grained than those of both senior and junior experts. However, in rare cases (3/118), human experts are better at handling open issues based on their knowledge and experience without much guidance.
\end{tcolorbox}

\subsection{Fixes are Non-trivial to Devise by Humans}

We further assess whether the 118 AC fixes attempted by \tool are non-trivial to devise by humans based on the results of the surveyed second question.
The results indicated that on average, junior and senior experts respectively identified 58.0 (49.15\%) and 53.5 (45.34\%) \textit{NO} cases that are not straightforward to propose a patch.
Based on majority voting, there were 42 (35.59\%) \textit{YES} cases in which most experts agreed on the straightforwardness of patch instrumentation. After manually inspecting them, we found they mostly ($88\%$) belong to \texttt{initialization} and \textit{changes of ownership}, which are straightforward to fix because only the \textit{owner of the contract} should be checked.
For the rest of the non-trivial cases, various role-permission pairs were included.
For instance, \texttt{Mint}, which could be subject to abundant rules to implement the AC policy, is not straightforward to fix.
Another example is the \texttt{Guardian} role~\cite{erc20tokenGuardian} for ERC20.
In this case, the original contract has three roles, \texttt{Guardian}, \texttt{Governor}, and \texttt{Minter}, which already form a hierarchy.
To provide a proper fix, the practitioner must understand the hierarchy and the functionality of the target function, which could be a laborious and complicated task.
These non-trivial cases could be effectively tackled by \name given that it has learned various rules and understood the hierarchical relationships of existing RBAC.

\noindent
\textbf{Case Study.}
We selected a case successfully repaired by \name, which was agreed by all participants to be tough to fix, to demonstrate \name's advantage.
The case is the motivating example in \myfig~\ref{lst:motivationexample}~\cite{example}.
The fix is complicated because one has to understand the implicit design of the \textit{external} function \texttt{depositFromOtherContract}.
We first explain the reason behind the correct patch: As there is already a \texttt{deposit} function to deposit one's own values belonging to \texttt{msg.sender}, the vulnerable function \texttt{depositFromOtherContract} takes in an argument \texttt{\_from} to deposit values on behalf of other users.
However, depositing on behalf of others requires a trustworthy authority to act as a centralized agency.
There are two trusted addresses defined in this contract, namely, \textit{owner} and \textit{bank}.
Given that \textit{bank} is set by \textit{owner}, the proper role within the context to manage deposits is \textit{bank}, which is exactly how the original author fixed it.

From the above case, we derive several challenges of manual patching:
(1) Going through existing functions and distinguishing them from each other regarding the desired functionality, i.e., \texttt{depositFromOtherContract} and \texttt{deposit};
(2) Understanding the existing RBAC hierarchy based on the implementation of the chain of trust, i.e., \textit{owner} and \textit{bank};
(3) Understanding the implicit relationship between the design of a centralized agency and the existing role \textit{bank}.
In response to these challenges, \tool could effectively mine the existing RBAC roles and implementations of the two existing deposit functions.
Then, GPT-4 could understand the implicit logic between them to address challenges (2) and (3).

\begin{tcolorbox}[boxrule=0.5pt,arc=1pt,boxsep=-1mm,breakable]
\textbf{Takeaway}:
Around half of the AC fixes are non-trivial to devise by humans, indicating that \name can provide a unique complement to assist human-in-the-loop repair as a copilot.
Through a case study, we conclude that with the aid of an LLM, the implicit logic can be dissected and streamlined from the source code, which is imperative for generating proper patches for AC vulnerabilities.
\end{tcolorbox}

%% file: rev_tex_2/7-discussion.tex
\section{Threats of Validity}
\rev{
\textbf{Internal Threats:}
The primary threat to \tool is the precision of static analysis.
As \tool mostly relies on Antlr to resolve dependency relationships of code statements using AST, rather than IR, \rcg may not achieve high precision and recall.
However, this potential inaccuracy does not markedly affect \name's capabilities for two reasons.
First, the selection of role-permission pairs primarily depends on GPT-4's logical reasoning capabilities, provided there is sufficient context to infer role and permission.
Second, in most cases, \tool performs static analysis within a single contract file.
This means that most of the call graphs, def-use chains of state variables, and mappings between parameters of functions could reliably rely on name mappings. Therefore, the static analysis in \tool may be flawed, but it suffices to support context understanding of GPT-4.}

\rev{An internal threat to validity stems from the dataset used for evaluating AC vulnerabilities, which was gathered primarily from limited online sources, specifically DefiHackLabs~\cite{defihacklabs} and tintinweb~\cite{tintinwebvul}, resulting in unequal representation and potential incompleteness. Such imbalanced distributions across data sources might inadvertently introduce biases. To mitigate this issue, we designed RQ4 explicitly to analyze the performance of \tool and baseline methods within individual categories, thereby reducing sensitivity to data imbalance. Additionally, we made considerable efforts to include as many cases as possible from diverse Internet sources, enlarging the dataset to facilitate a more comprehensive and fair comparison.
}

\rev{The last external threat to validity lies in the interpretability of LLMs like GPT-4, which \tool relies on. Due to their black-box nature, LLMs may generate outputs that are difficult to explain or verify, potentially introducing incorrect or inconsistent repairs. Issues such as hallucination, prompt sensitivity, and lack of transparency in reasoning pose risks, particularly in security-critical contexts like access control. To mitigate these concerns, we constrain the LLM’s output space using a dynamic taxonomy of RBAC role-permission pairs, reducing the likelihood of invalid predictions. Additionally, we integrate static analysis checks to validate the syntactic and semantic correctness of generated patches. We also employ carefully crafted in-context prompts to enhance stability and reduce variation across similar cases. Furthermore, \tool incorporates a dual-agent feedback framework, where a validation agent assesses the generated output and provides feedback to the generator, enabling iterative refinement. Together, these mechanisms help enhance interpretability and reliability, though we acknowledge that challenges inherent to LLMs remain an open research issue.}

\rev{
\textbf{External Threat:}
The potential threat to validity arises from the initial reliance on a manually curated RBAC taxonomy. Due to inherent constraints in the available dataset, this taxonomy may not comprehensively represent all possible role-permission relationships encountered in real-world smart contract implementations. Such incompleteness could potentially lead to inaccuracies or misidentifications of role-permission pairs during vulnerability repair. To address this limitation, we incorporated an adaptive mechanism within our approach, enabling the automatic addition of newly identified role-permission pairs to dynamically expand the taxonomy. This strategy effectively mitigates the risks posed by a static, finite taxonomy, ensuring greater robustness and adaptability of the proposed solution.
}

%% file: rev_tex_2/8-relatedwork.tex
\section{Related Work}

\noindent
\subsection{Smart Contract Repair}
Smart contract vulnerability repair has seen significant progress, such as Aroc~\cite{Jin2022aroc}, SmartShield~\cite{zhang2020smartshield}, SGuard~\cite{nguyen2021sguard}, Elysium~\cite{ferreira2022elysium}, SCRepair~\cite{yu2020smart}, and SmartFix~\cite{so2023smartfix}. However, research on repairing AC-related vulnerabilities remains limited. Tools like Aroc and SmartShield do not support AC repairs, SGuard addresses only \textit{tx.origin} misuse, and Elysium fixes only two sensitive operations. SCRepair's effectiveness is constrained by manual unit tests, while SmartFix handles only \textit{tx.origin} and \textit{re-initialization} vulnerabilities.

In light of the above, \tool stands out in two ways: 
\ding{172} Human-Level Reasoning: We address and resolve the limitations inherent in prior works that relied solely on predefined templates.
By utilizing GPT-4, our method engages in conversational sessions employing CoT and MAD, which allows \tool to achieve human-like reasoning. This marks a significant advancement in the methodology for AC vulnerability repairs.
\ding{173} Comprehensive Coverage: Many existing tools support AC vulnerabilities but are often restricted to handling conventional patterns.
In contrast, \tool boasts the capability to address AC vulnerabilities across diverse scenarios.

\noindent
\subsection{Traditional Program Repair}
Numerous works have focused on repairing bugs or vulnerabilities in traditional software~\cite{son2013fix,xia2023automated,xia2023revisiting,xia2023keep, wei2023copiloting, fu2022vulrepair,chi2022seqtrans,chen2022neural,jiang2018shaping,kim2013automatic,long2015staged,long2016automatic,weimer2009automatically}, especially in C~\cite{fu2022vulrepair,chi2022seqtrans,chen2022neural}, Java~\cite{jiang2018shaping,wei2023copiloting}, and PHP~\cite{son2013fix}.
Moreover, several concurrent works~\cite{xia2023automated,xia2023revisiting,xia2023keep,wei2023copiloting,FineTuneRepair23,ImpactLMforRepair23} propose LLM-based APR solutions for bug fixes.
For example, Xia et al.~\cite{xia2023automated} studied the effectiveness of LLMs for APR and found that LLMs generally outperform traditional approaches.
ChatRepair~\cite{xia2023keep} employs multiple sessions for interactive repair with GPT-4.
Repilot~\cite{wei2023copiloting} innovatively combines the completion engine with LLM to synergistically generate patches.
FitRepair~\cite{xia2023revisiting} leverages the \textit{plastic surgery hypothesis} to repair bugs using existing code ingredients by performing static analysis and information retrieval on the source code.
Other related works~\cite{fu2022vulrepair,chi2022seqtrans,chen2022neural,jiang2018shaping,kim2013automatic,long2015staged,long2016automatic,weimer2009automatically} mostly employ traditional methods, such as Neural Machine Translation, to synthesize repairs for bugs or iteratively search for proper patches.
Our work shares several common practices, such as conversational sessions and existing ingredient reuse, but uniquely mines RBAC practices and relevant code context to guide LLMs.

%% file: rev_tex_2/9-future_works.tex
\section{Future Work}
\rev{
Building upon the insights and infrastructure established by \tool, we identify several promising directions for future exploration:
\begin{itemize}[leftmargin=5pt]
\item \textbf{LLM-guided Repair of Advanced Vulnerabilities.} While this work focuses on AC vulnerabilities, our approach establishes a foundation for addressing a broader class of complex smart contract vulnerabilities. In future work, we plan to extend \tool's reasoning and repair capabilities to additional vulnerability types that require deep semantic understanding, such as reentrancy with indirect triggers, improper state transitions, delegatecall misuse, and incorrect payment logic. These categories often involve non-trivial control flow, cross-contract dependencies, or subtle logic flaws that static patterns alone cannot effectively capture. Enhancing \tool with formal specifications, symbolic reasoning, or integration with domain-specific ontologies may further improve its adaptability and accuracy. 
\item \textbf{Detection and Repair of Multi-Function AC Vulnerabilities.} A more advanced but rare class of AC vulnerabilities involves multiple functions collectively contributing to unauthorized privilege escalation. These vulnerabilities are especially challenging, as they may not exhibit direct or transitive call relationships but instead share critical state variables that facilitate cross-function interactions. We intend to investigate this class of vulnerabilities by modeling state-dependent attack surfaces and designing analysis techniques to identify latent privilege escalation paths that span disjointed code regions. This form of vulnerability is rare but significantly harder to detect and mitigate.
\item \textbf{Adaptive Taxonomy Evolution via Online Learning.}
While \tool leverages a dynamic taxonomy mined from a large corpus of on-chain contracts, smart contract development practices continue to evolve, introducing new roles, patterns, and access semantics. This new emerging knowledge may not be accommodated well by merely updating the taxonomy with more RBAC pairs. To maintain robustness and adaptability, we envision extending the current static RBAC taxonomy into a dynamic, Retrieval-Augmented Generation (RAG) system. In this setting, the LLM would query an updatable knowledge base of role-permission pairs—continuously refined from emerging contracts, validator feedback, and user interaction logs—enabling it to incorporate the latest access control practices during repair. This dynamic integration would reduce reliance on static assumptions, enhance coverage of long-tail or novel RBAC cases, and provide a pathway for \tool to generalize beyond its original training distribution with minimal human intervention.
\item \textbf{LLM-guided AC Vulnerability Detection.} With the mined RBAC taxonomy and enhanced context comprehension mechanisms, we plan to extend the scope of \tool from repair to detection. By leveraging LLMs’ ability to semantically understand AC context, although ACFix has advanced the usage of LLM on smart contract security repair, we aim to further detect improper RBAC implementation by identifying role-permission mismatches relative to the intended functionality of contracts. However, vulnerability detection presents unique challenges compared to repair, particularly in terms of scalability. Unlike repair, which starts from known vulnerable functions, detection must assume that any function could be misconfigured and thus requires comprehensive analysis across the entire contract. This substantially increases computational costs, as LLMs must perform intricate RBAC reasoning for all functions. To address this, we plan to incorporate static filtering techniques to preselect likely-vulnerable candidates, enabling scalable and efficient LLM-guided detection without exhaustive analysis.
\item \minor{\textbf{Advanced Validation Paradigm.} We plan to further advance the validation process by exploring more sophisticated MAD frameworks. In particular, we aim to incorporate additional specialized agents, such as adversarial critics and domain-specific oracles, to enrich the debate dynamics and improve the reliability of patch validation. We also intend to investigate adaptive debate strategies, where the validation process dynamically adjusts the roles or number of agents based on the complexity of the repair task. These directions are expected to enhance both the robustness and explainability of the validation stage, and will be integrated into future iterations of \tool.}
\end{itemize}
}

%% file: rev_tex_2/10-conclusion.tex
\section{Conclusion}

This paper proposed \tool for repairing AC vulnerabilities in smart contracts by guiding LLMs with AC practices and code context. We developed an RBAC taxonomy from on-chain contracts and a slicing algorithm to extract AC-related context. Equipped with check rules and \val of MAD, \tool repaired 94.92\% of cases in our dataset, outperforming existing tools. Our evaluation included
 a human study to assess the quality of \tool's repairs compared to humans'.

 \section*{Acknowledgment}
This research is supported by the Ministry of Education, Singapore, under its Academic Research Fund Tier 1 (RG96/23). It is also supported by the National Research Foundation, Singapore, and DSO National Laboratories under the AI Singapore Programme (AISG Award No: AISG2-GC-2023-008); by the National Research Foundation Singapore and the Cyber Security Agency under the National Cybersecurity R\&D Programme (NCRP25-P04-TAICeN); and by the Prime Minister’s Office, Singapore under the Campus for Research Excellence and Technological Enterprise (CREATE) Programme.
Any opinions, findings and conclusions, or recommendations expressed in these materials are those of the author(s) and do not reflect the views of the National Research Foundation, Singapore, Cyber Security Agency of Singapore, Singapore.

\section*{Open Science Policy}
To facilitate replication and future research, we have released our source code and dataset on an anonymous website~\cite{dataset}.